\documentclass[twocolumn,superscriptaddress,nobibnotes,nofootinbib,floatfix,citeautoscript,aip,jcp,reprint]{revtex4-1}

\pdfoutput=1

\usepackage{graphicx} 
\usepackage{amssymb,amsmath} 
\usepackage{verbatim} 
\usepackage{mathrsfs}

\begin{document}

\newcommand{\sanseb}{University of the Basque Country UPV/EHU, Nano-Bio Spectroscopy Group, Avenida de Tolosa 72, 20018 San Sebastian, Spain}
\newcommand{\lpmcn}{Universit\'e de Lyon, F-69000 Lyon, France and
LPMCN, CNRS, UMR 5586, Universit\'e Lyon 1, F-69622 Villeurbanne, France}
\newcommand{\rome}{Department of Physics, University of Rome "La Sapienza", Piazzale Aldo Moro 5, 00185 Roma, Italy}
\newcommand{\fhi}{ Fritz-Haber-Institut der Max-Planck-Gesellschaft, Faradayweg 4-6, D-14195 Berlin, Germany}
\newcommand{\halle}{Max-Planck Institute for Microstructure Physics, Weinberg 2, 06120 Halle, Germany}

\title{Ab-initio angle and energy resolved photoelectron spectroscopy with time-dependent density-functional theory} 
\date{\today} 
\author{U. De Giovannini} \email{umberto.degiovannini@ehu.es} \affiliation{\sanseb} 
\author{D. Varsano}, \affiliation{\rome} 
\author{M. A. L. Marques} \affiliation{\lpmcn} 
\author{H. Appel} \affiliation{\fhi} 
\author{E. K. U. Gross} \affiliation{\halle} 
\author{A. Rubio} \affiliation{\sanseb}

%
%
\begin{abstract}
We present a time-dependent density-functional method able to
describe the photoelectron spectrum of atoms and molecules when
excited by laser pulses. This computationally feasible scheme is
based on a geometrical partitioning that efficiently gives access to
photoelectron spectroscopy in time-dependent density-functional calculations. By
using a geometrical approach, we provide a simple description of
momentum-resolved photoemission including multi-photon effects.  The
approach is validated by comparison with results in the
literature and exact calculations. Furthermore, we present numerical photoelectron angular
distributions for randomly oriented nitrogen molecules in a short near
infrared intense laser pulse and helium-(I) angular spectra for aligned
carbon monoxide and benzene.
\end{abstract}
\maketitle

%
%
\section{Introduction} \label{sec:intro}

Photoelectron spectroscopy is a widely used technique to analyze the
electronic structure of complex
systems~\cite{Ellis:2005te,Wu:2011cv}. The advent of intense
ultra-short laser sources has extended the range of applicability of
this technique to a vast variety of non-linear phenomena like 
high-harmonic generation, above-threshold ionization (ATI), bond softening
and vibrational population trapping~\cite{Brabec:2000iz}. Furthermore,
it turned attosecond time-resolved pump-probe photoelectron
spectroscopy into a powerful technique for the characterization of
excited-states dynamics in nano-structures and biological
systems~\cite{Krausz:2009hz}.  Angular-resolved ultraviolet
photoelectron spectroscopy is by now established as a powerful
technique for studying geometrical and electronic properties of
organic thin films~\cite{Puschnig:2009ho,Kera:2006fv}.  Time-resolved
information from streaking spectrograms~\cite{Schultze:2010kr},
shearing interferograms~\cite{Remetter:2006fc}, photoelectron
diffraction~\cite{Meckel:2008vn}, photoelectron
holography~\cite{Huismans:2011kh}, etc. hold the promise of
wavefunction reconstruction together with the ability to follow the
ultrafast dynamics of electronic wave-packets. Clearly, to complement
all these experimental advances, and to help to interpret and
understand the wealth of new data, there is the need for ab-initio
theories able to provide (time-resolved) photoelectron spectra (PES)
and photoelectron angular distributions (PAD) for increasingly complex
atomic and molecular systems subject to arbitrary perturbations (laser intensity and shape).

Photoelectron spectroscopy is a general term which refers to all
experimental techniques based on the photoelectric effect.  In photoemission
experiments a light beam is focused on a sample, transferring energy to
the electrons. For low light intensities an electron can absorb a
single photon and escape from the sample with a maximum kinetic energy
$\hbar\omega -I_P$ (where $\omega$ is the photon angular frequency and $I_P$ the
first ionization potential of the system) while for high intensities electron
dynamics can be interpreted considering a three-step
model~\cite{Corkum:1993ee}. This model provides a semiclassical
picture in terms of ionization followed by free electron propagation
in the laser field with return to the parent ion, and
rescattering. Such rescattering processes are the source of many
interesting physical phenomena. In the case of long pulses, for
instance, multiple photons can be absorbed resulting in emerging
kinetic energies of $s \hbar\omega -I_P -U_P$ (where $s$ is the
number of photons absorbed, $U_P=\epsilon^{2}/4\omega^{2}$ is the ponderomotive 
energy and $\epsilon$ the electric field amplitude) 
forming the so called ATI peaks in the resulting photoelectron spectrum. 
In all cases the observable is the
escaping electron momentum measured at the detector.

In general, the interaction between electrons in an atom or molecule
and a laser field is difficult to treat theoretically, and several
approximations are usually performed.  Clearly, a full many-body
description of PES is prohibitive, except for the case of few (one or
two) electron
systems~\cite{Lein:2001fj,Pazourek:2011wi,Ishikawa:2012gx}.  As a
consequence, the direct solution of the time dependent Schr\"odinger
equation (TDSE) in the so-called single-active electron (SAE)
approximation is a standard investigation tool for many strong-field
effects in atoms and dimers and represents the benchmark for analytic
and semi-analytic models~\cite{Schafer:1990gga,Chelkowski:1998ec,
  Grobe:1999vy,Chen:2006fa,Tong:2006wv,
  Tong:2007vd,Awasthi:2008hg,Petretti:2010vh,Schultze:2010kr,Burenkov:2010js,Huismans:2011kh,Zhou:2011jm,Tao:2012ev,Catoire:2012wk}.
Perturbative approaches based on the standard Fermi golden rule are
usually employed. For weak lasers, plane wave methods~\cite{Puschnig:2009ho} 
and the independent atomic center
approximation~\cite{Grobman:1978jd} have been applied, while in the strong
field regime, Floquet theory, the strong-field
approximation~\cite{Huismans:2011kh,GazibegovicBusuladzic:2011cp} and
semiclassical methods~\cite{Corkum:1993ee, Lewenstein:1995ca,
  Paulus:1999ks} are routinely used.

From a numerical point of view, it would be highly desirable to have a
PES theory based on time-dependent density functional theory (TDDFT)~\cite{Runge:1984ig,Marques:2011ud}
where the complex many-body problem is described in terms of a fictitious 
single-electron system.  
For a given initial many body state, TDDFT maps the
whole many-body problem into the time dependence of the density from
which all physical properties can be obtained. The method is in principle
exact, but in practice approximations have to be made for the
unknown exchange-correlation functional as well as for specific
density-functionals providing physical observables. This latter issue
is much less studied than the former, and to the best of our knowledge
a formal derivation of momentum-resolved PES from the time dependent
density has not been performed up to now.  In any case, several works
were published addressing the problem of single and multiple
ionization processes within TDDFT. For example, ionization rates were
calculated for atoms and
molecules~\cite{Chu:2004ki,Telnov:2009eb,Son:2009kx,Son:2009iz,Hansen:2011gz},
and TDDFT with the sampling point method (SPM) has been employed in
the study of PES and PAD for sodium
clusters~\cite{Pohl:2000ut,Pohl:2004ef,Wopperer:2010hj,Wopperer:2010fn}.

In this work, besides presenting a formal derivation of a photoelectron orbital
functional, we report on a new and physically sound scheme to compute PES of
interacting electronic systems in terms of the time-dependent single
electron Kohn-Sham (KS) wavefunctions.  The scheme relies on geometrical considerations and 
is based on a splitting technique~\cite{Chelkowski:1998ec,
  Grobe:1999vy,Chen:2006fa,Tong:2006wv, Tong:2007vd}. The idea is based on the
of partitioning of space in two regions (see Fig.~\ref{fig:geometrical_scheme} below): 
In the inner region, the KS wave function is obtained by solving the TDDFT equations numerically; in
the outer region, electrons are considered as free particles, 
the Coulomb interaction is neglected, and the wavefunction 
is propagated analytically with only the laser field.
Electrons flowing from the inner region to the outer region are
recorded and coherently summed up to give the final result.  In
addition to the adaptation of the traditional splitting procedure to
TDDFT, we propose a novel scheme where electrons can seamlessly drift
from one region to the other and spurious reflections are greatly
suppressed.  This procedure allows us to reduce considerably the
spatial extent of the simulation box without damaging the accuracy of the method.

The rest of this Article is organized as follows. The formalism for
describing photoelectrons in TDDFT is delineated in
Sect.~\ref{sec:theory}.  In order to make contact with the literature,
we first give a brief introduction to the state-of-the-art for the
ab-initio calculation of PES for atomic and molecular systems. In
Sect.~\ref{sub:phase-space_interpretation} we introduce the
geometrical approach in the context of quantum phase-space. The
phase-space approach is then derived in the case of effective
single-particle theories like TDDFT in
Sect.~\ref{sub:phase_space_tddft}. In Sect.~\ref{sub:the_mask_method}
we introduce the mask method, an efficient propagation scheme based on
space partitioning.

Three applications of the mask method are presented in
Sect.~\ref{sec:applications}.  One application deals with the hydrogen
atom and illustrates the different mask methods in a simple
one-dimensional model also in comparison with the sampling point
method~\cite{Pohl:2000ut}. The above threshold ionization of
three-dimensional hydrogen is examined and compared with values from
the literature.  In the second application we illustrate PADs from
randomly oriented nitrogen molecules in a strong near-infrared
ultra-short laser pulse. Comparison with the experiment and molecular
strong-field approximation is
discussed~\cite{GazibegovicBusuladzic:2011cp}.  The third application
of the method regards helium-(I) (wavelength 58 nm) PADs for oriented carbon monoxide and
benzene. Results are discussed in comparison with the plane wave
approximation.  Finally, in Sect.~\ref{sec:conclusions} we discuss the
results and present the conclusions.

All our numerical calculations were performed with the real-time,
real-space TDDFT code \texttt{Octopus}~\cite{Castro:2006ve,Andrade:2012}, freely
available under the GNU public license. Atomic units are used
throughout unless otherwise indicated.

%
%
\section{Modeling photoelectron spectra} 
\label{sec:theory}

In order to put in perspective the results of the present Article, we
will give a brief introduction on the status of the principal
techniques available for ab-initio PES calculations. We start our
description with the methods employed to study one-electron systems.

For one-electron systems PES can be calculated exactly from the direct
solution of the TDSE. Several methods have been employed to extract PES
information from the solution of the TDSE.  The most direct and
intuitive way is via direct projection methods where the PES is
obtained by projecting the wave function at the end of the pulse onto
the eigenstates describing the continuum. These eigenstates are
extracted through the direct diagonalization of the Hamiltonian
without including the interaction with the field. The momentum
probability distribution can then be easily obtained from the Fourier
transform of the continuum part of the time-dependent
wavefunction\cite{Burenkov:2010js}.

Another approach, that avoids the calculation of the full continuum
spectrum, involves the analysis of the exact wavefunction
$|\Psi\rangle$ after the laser pulse via a resolvent
technique~\cite{Schafer:1990gga,Catoire:2012wk}. In this case, the
energy resolved PES is given by the direct projection on out-going
wavefunctions with $ P(E)=|\langle \Phi(E)|\Psi\rangle|^2=
\langle\Psi|\hat{D}(E)|\Psi\rangle $, where $\Phi(E)$ denotes an
out-going (unbound) electron of energy $E$ of the laser-free
Hamiltonian, and $\hat{D}(E)$ is the corresponding projection operator
that can be conveniently
approximated~\cite{Schafer:1990gga,Catoire:2012wk}.

Normally, one needs accurate wave functions in a large space domain to
obtain the correct distribution of the ejected electrons. This is
because the unbound parts of the wave packet spread out of the core
region, and conventional expressions for the transition amplitude
need these parts of the wave function. Solving the TDSE within all
the required volume in space can easily become a very difficult
computational problem. Several techniques were developed during the
years to solve the problem. For simple cases these difficulties can be
overcome by the use of spherical coordinates. Geometrical splitting
techniques have also been employed~\cite{Chelkowski:1998ec,
  Grobe:1999vy,Chen:2006fa,Tong:2006wv, Tong:2007vd}. Furthermore,
formulations in the Kramers-Henneberger frame of
reference~\cite{Telnov:2009ie} and in
momentum-space~\cite{Zhou:2011jm} led to calculations with
remarkable high precision. Recently a promising surface flux method
has also been proposed~\cite{Tao:2012ev}.

The exact solution of the TDSE in three dimensions for more than two electrons is unfeasible
and the limit rises to four electrons for one-dimensional models~\cite{Helbig:2011bx}. Due to this limitation 
basically all ab-initio calculations for
multi-electron systems are preformed under the SAE approximation. In
the SAE only one electron interacts with the external field while the
other electrons are frozen~\cite{Awasthi:2008hg}, and the TDSE is thus
solved only for the active electron. This approximation has been
successfully employed in several photoemission studies for atoms and
molecules in strong laser
fields~\cite{Petretti:2010vh,Schultze:2010kr,Huismans:2011kh}. However,
the failure of this simple model to describe multi-electron (correlation) effects
calls for better schemes~\cite{Petretti:2010vh}.

The inclusion of exchange-correlation effects for a system of many
interacting electrons can be achieved within TDDFT while keeping the
simplicity of working with a set of time-dependent (fictitious)
single-particle orbitals.  In spite of transferring all the many-body
problem into an unknown exchange-correlation functional, the lack of a
density functional providing the electron emission probability is a
major limitation for a direct access to photoelectron observables from
the time evolution of the density (note that, in spite of the Runge-Gross 
theorem~\cite{Runge:1984ig} stating that all observables are functionals of the time
dependent density, in practice we know few observables that can 
be written in terms of the time-dependent density, one example being the absorption spectra).

There has been some attempts to describe PES and multiple ionization
processes with TDDFT in the standard adiabatic
approximation~\cite{Chu:2004ki,Telnov:2009eb,Son:2009kx,Son:2009iz,Hansen:2011gz}. All
these works use boundary absorbers to separate the bound and continuum
part of the many-body wavefunction. The emission probability is then
correlated with the time dependence of the number of bound electrons.

An alternative and simple scheme is provided by the
SPM~\cite{Pohl:2000ut}. Here the idea is to record single-particle 
wavefunctions in time at a fixed sampling point $r_{S}$ away from
the core. The time Fourier transform of the wavefunction recorded at
$r_{S}$ represents the probability of having an electron in $r_{S}$
with energy $E$. The probability to detect one electron with energy
$E$ in $r_{S}$ is then given by the sum over all occupied orbitals:
\begin{equation}\label{eq:sampling_point}
	P_{{\bf r}_{S}}(E)=\sum_{i=1}^{occ.}|\psi_i({\bf r}_{S},E)|^2 \,.
\end{equation}
This method is easy to implement, can be extended to give also angular
information~\cite{Wopperer:2010fn}, and is also clearly applicable to
the TDSE in the SAE.  However, it lacks formal derivation as it is
directly based on Kohn-Sham wavefunctions without a direct connection
to the many-body state. Furthermore, it is strongly dependent on the
position of the sampling point and the minimum distance. This distance
sometimes turns out to be quite large in order to avoid artifacts, and
is strongly dependent on the laser pulse properties. We discuss
further details concerning this method in
Sect.~\ref{sub:multi_photon_ionization_1d}.

In the following we present an alternative method inspired by
geometrical splitting and derive it from a phase-space point of view.
The method can be naturally converged by increasing the size 
of the different simulation boxes.

%
%
\subsection{Phase-space geometrical interpretation}
\label{sub:phase-space_interpretation}

An intuitive description of photoelectron experiments can be obtained
resorting to a phase-space picture.  Experimental detectors are able
to measure photoelectron velocity with a certain angular distribution
for a sequence of ionization processes with similar initial
conditions.  The quantity available at the detector is therefore
connected to the probability to register an electron with a given
momentum ${\bf p}$ at a certain position ${\bf r}$. From this
consideration it would be tempting to interpret photoemission
experiments with a joint probability distribution in the phase-space
$({\bf r},{\bf p})$. Such a classical picture however conflicts with the
fundamental quantum mechanics notion of the impossibility to
simultaneously measure momentum and position, and prevents us from
proceeding in this direction. A link between the classical and quantum
picture is needed beforehand.

In order to make a connection to a microscopic description it turns
out to be convenient to extend the classical concept of phase-space
distributions to the quantum realm.  A common prescription comes from
the Wigner transform of the one-body density matrix with respect to
the center of mass ${\bf R}=({\bf r}+{\bf r}^\prime)/2$ and relative
${\bf s}= {\bf r}-{\bf r}^\prime$ coordinates. The $d$-dimensional
(here and after $d\leq 3$) transform is defined as
\begin{equation}\label{eq:wigner}
	w({\bf R},{\bf p},t)=\int \frac{{\rm d} {\bf s}}{(2\pi)^{\frac{d}{2}}}\, e^{i {\bf p} \cdot {\bf s}} \rho({\bf R}+{\bf s}/2,{\bf R}-{\bf s}/2,t)\,, 
\end{equation}
with 
\begin{multline}\label{eq:onedensity}
      \rho({\bf r},{\bf r}^\prime,t)=\int {\rm d}{\bf r}_2\dots{\rm d}{\bf r}_N \Psi({\bf r},{\bf r}_2,\dots,{\bf r}_N,t)  \\ 
      \times\Psi^{*}({\bf r}^\prime,{\bf r}_2,\dots,{\bf r}_N,t)\,, 
\end{multline}
being the one-body density matrix, and $\Psi({\bf r}_1,{\bf
  r}_2,\dots,{\bf r}_N,t)$ the $N$-body wavefunction of the system at
time $t$. The Wigner function defined above is normalized and its
integral over the whole space (momentum) gives the probability to find
an electron with momentum ${\bf p}$ (position ${\bf R}$). As the
uncertainty principle prevents the simultaneous knowledge of position
and momentum, $w({\bf R},{\bf p})$ cannot be a proper joint
distribution. Moreover it can assume negative values due to
nonclassical dynamics. Nevertheless the Wigner function $w({\bf
  R},{\bf p})$ constitutes a concept close to a probability
distribution in phase space $({\bf R},{\bf p})$ compatible with
quantum mechanics.
\begin{figure}
  \includegraphics[width=8.6cm,keepaspectratio]{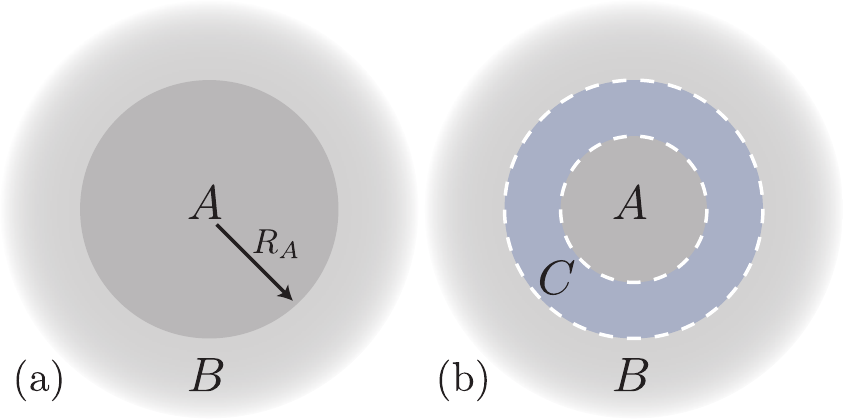}
  \caption{(Color online) Schematic description of (a) the partitioning of space for the
    phase space method and (b) the mask method. Region $A$ is the
    interaction region, $B$ is the Volkov propagation region and $C$
    is the overlap region where $\Psi_A$ and $\Psi_B$ mix under the
    mask function.}
  \label{fig:geometrical_scheme}
\end{figure}

The quantum phase-space naturally leads to a geometrical
interpretation of photoemission.  One could think to divide the space
in two regions $A$ and $B$ as in
Fig.~\ref{fig:geometrical_scheme}~(a), where region $B$ represents the
region where detectors are positioned and $A$ is defined as the
complement of $B$. In this picture, PES can be seen as the probability
to have an electron with given momentum in $B$.  It is then natural to
define the momentum-resolved photoelectron spectrum as
\begin{equation}\label{eq:pes_wigner}
  \mathcal{P}(\mathbf{p})=\lim_{t\rightarrow \infty}\int_{B} \mathtt{d}\mathbf{R}\, w({\bf R},{\bf p},t)
\end{equation}
where the spatial integration is carried out in region $B$, and the
limit $t\rightarrow \infty$ assures that region $B$ contains all
photoelectron contributions. From the knowledge of the
momentum-resolved PES [cf. Eq.~(\ref{eq:pes_wigner})] one can access
several different quantities by simple integration. For instance, in
three dimensions ($d=3$) the energy-resolved PES is obtained
integrating over the solid angle $\Omega$:
\begin{equation}\label{eq:pes_tddft_e}
  P(E=p^{2}/2)=\int_{0}^{4\pi} {\rm d}\Omega\, \mathcal{P}(\mathbf{p})\, ,
\end{equation}
and the photoelectron angular distribution in the system reference frame is given by,
\begin{equation}\label{eq:pad}
  P(E=p^{2}/2,\theta)=\int_{0}^{2\pi} {\rm d} \phi\, \mathcal{P}(\mathbf{p})\,.
\end{equation}

In spite of giving an intuitive picture of PES,
Eq.~(\ref{eq:pes_wigner}) is not suited for direct numerical
evaluation since it requires the knowledge of the full one-body 
density matrix in the whole space. In the next
section we will make a contact with effective single particle theories
like TDDFT to overcome the limitations due to the knowledge of the many-body 
wavefunction. In order to avoid integration over the whole space 
an efficient evolution scheme is presented in Sect.~\ref{sub:the_mask_method}.


\subsection{Phase space interpretation within TDDFT } 
\label{sub:phase_space_tddft}

TDDFT is an effective single particle theory where the many-body
wavefunction is described by an auxiliary single Slater determinant
$\Psi_{KS}(\mathbf{r_1},\dots,\mathbf{r_N})$ built out of Kohn-Sham
orbitals $\psi_i(\mathbf{r})$ \cite{Runge:1984ig,Marques:2011ud}.  In
order to simplify the notation, we drop the explicit time dependence from
the wavefunctions and assume that the following equations are written
in the limit $t\rightarrow \infty$ as prescribed by
Eq.~(\ref{eq:pes_wigner}).

Being represented by a single determinant, the one-body Kohn-Sham
density matrix is given by
\begin{equation}
  \rho_{KS}(\mathbf{r},\mathbf{r}^{\prime})=\sum_{i=1}^{\rm occ.} \psi_i(\mathbf{r})\psi_i(\mathbf{r}^{\prime})
\end{equation}
where the sum in carried out over all occupied orbitals.  Performing a
decomposition of each orbital according to the partition of
Fig.~\ref{fig:geometrical_scheme}~(a) we obtain
\begin{equation}\label{eq:orbital_partition}
  \psi_{i}(\mathbf{r})=\psi_{A,i}(\mathbf{r})+\psi_{B,i}(\mathbf{r})\, , 
\end{equation} 
where $\Psi_{A,i}(\mathbf{r})$ is the part of the wavefunction
describing states localized in $A$ and $\Psi_{B,i}(\mathbf{r})$ is the
ionized contribution measured at the detector in $B$. The one-body
density matrix can now be accordingly decomposed as a sum of four terms
\begin{multline}\label{eq:one_dens_TDDFT}
  \rho_{KS}(\mathbf{r},\mathbf{r}^{\prime})=\sum_{i=1}^{\rm occ.} \left[   \psi_{A,i}(\mathbf{r})\psi_{A,i}^{*}(\mathbf{r}^{\prime}) +    \psi_{A,i}(\mathbf{r})\psi_{B,i}^{*}(\mathbf{r}^{\prime}) \right.  \\
    +\left.   \psi_{B,i}(\mathbf{r})\psi_{A,i}^{*}(\mathbf{r}^{\prime}) + \psi_{B,i}(\mathbf{r})\psi_{B,i}^{*}(\mathbf{r}^{\prime}) \right].
\end{multline}
From Eq.~(\ref{eq:one_dens_TDDFT}) we can build the KS Wigner function
defined in Eq.~(\ref{eq:wigner}) and obtain the momentum-resolved
probability distribution by inserting it into
Eq.~(\ref{eq:pes_wigner}). We note that this step involves a
non-trivial approximation, namely that the KS one-body density matrix
is a good approximation to the fully interacting one in region B. This
is, however, much milder than the assumption that the Kohn-Sham determinant 
is a good approximation to the many-body wavefunction in region B, as it is
done, e.g., in the SPM.

The final result is a sum of four overlap double integrals that can be
simplified further. For a detailed calculation we refer to
Appendix~\ref{sec:overlap_integral}.  The first overlap integral,
containing a product of two functions localized in $A$
[cf. Eq.~(\ref{eq:one_dens_TDDFT})], is zero due to the spatial
integration in $B$. The two next overlap integrals, containing mixed
products of wavefunctions localized in $A$ and $B$, can be reduced 
by increasing the size of region $A$. Assuming $A$ to be large
enough to render these terms negligible the only integral we are left
with is the one containing functions in $B$, leading to
\begin{multline}\label{eq:pes_wigner_tddft_BB}
\mathcal{P}(\mathbf{p})\approx\int_B{\rm d}\mathbf{R}\,\int \frac{{\rm d}\mathbf{s}}{(2\pi)^{\frac{d}{2}}}\, e^{i \mathbf{p}\cdot \mathbf{s}}
  \\
  \sum_{i=1}^{\rm occ.} \psi_{B,i}(\mathbf{R}+\frac{\mathbf{s}}{2})\psi_{B,i}^*(\mathbf{R}-\frac{\mathbf{s}}{2})\,.
\end{multline} 
The approximation sign $\approx$ is a reminder for the error committed
in discarding the mixed overlap integrals.  Since the probability of
finding an ionized electron in region $A$ is zero for $t\rightarrow
\infty$, we can extend the integration over $B$ in
Eq.~(\ref{eq:pes_wigner_tddft_BB}) to the whole space. Using the
integral properties of the Wigner transform we finally obtain
\begin{equation}\label{eq:pes_tddft}
  \mathcal{P}(\mathbf{p})\approx\sum_{i=1}^{\rm occ.} \left| \tilde{\psi}_{B,i}(\mathbf{p})\right|^2\,,
\end{equation} 
where $\tilde{\psi}_{B,i}(\mathbf{p})$ is the Fourier transform of
$\psi_{B,i}(\mathbf{r})$ and the expression is written in the limit
for $t\rightarrow \infty$.  Equation~(\ref{eq:pes_tddft}) gives an
intuitive formulation of momentum-resolved PES as a sum of the Fourier
component of each orbital in the detector region. It is worth to note
that Eq.~(\ref{eq:pes_tddft}) is not restricted to TDDFT and can be
applied to other effective single-particle formulations such as
time-dependent Hartree-Fock and the TDSE in the SAE approximation.

The numerical evaluation of the ionization probability from
Eq.~(\ref{eq:pes_tddft}) requires the knowledge of the wavefunction
after the external field has been switched off. For ionization
processes this means that one has to deal with simulation boxes that
extend over several hundred atomic units and this practically constrains
the method only to one-dimensional calculations. In the next section
we will derive a simple scheme to overcome this limitation making the present
scheme applicable for realistic simulations of molecules and nanostructures.


\subsection{The mask method} 
\label{sub:the_mask_method}

In the previous sections we described a practical way to evaluate the
momentum-resolved PES following the spatial partitioning of
Fig.~\ref{fig:geometrical_scheme}~(a) and how this can be conveniently
cast in the language of TDDFT.  In this section we take a step further
in developing an efficient time evolution scheme by exploiting the
geometry of the problem together with some physical assumptions.

\begin{figure}
  \includegraphics[width=8.6cm,keepaspectratio]{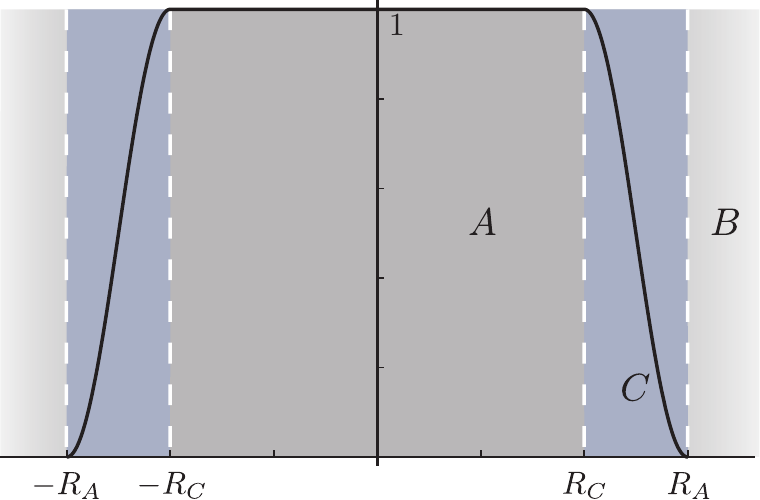}
  \caption{(Color online) The mask function implements a smooth transition from
    region $A$ to region $B$. The functional shape used for actual
    calculations is defined in Eq.~(\ref{eq:mask_function}).}
  \label{fig:mask_function}
\end{figure}

We start by introducing a split-evolution scheme: At each time $t$ we
implement a spatial partitioning of Eq.~(\ref{eq:orbital_partition})
as following
\begin{equation}\label{eq:orbital_partition_mask}  
  \left\{ 
  \begin{array}{ll}
    \psi_{A,i}(\textbf{r},t)=M(\textbf{r})\psi_{i}(\textbf{r},t) \\ 
    \psi_{B,i}(\textbf{r},t)=[1-M(\textbf{r})]\psi_{i}(\textbf{r},t) 
\end{array} \right. \,,
\end{equation}
where $M(\mathbf{r})$ is a smooth mask function defined to be 1 deep
in the interior of region $A$ and 0 outside, as shown in
Fig.~\ref{fig:mask_function}.
Such a mask function, along with the partitions $A$ and $B$, introduces
a buffer region $C$ (technically handled as the outermost shell of
$A$), where $\psi_{A,i}(\textbf{r},t)$ and $\psi_{B,i}(\textbf{r},t)$
overlap [see Fig.~\ref{fig:geometrical_scheme}~(b)].

We can set up a propagation scheme from time $t$ to $t^{\prime}$ as following 
\begin{equation}\label{eq:prop_scheme}  
  \left\{ 
  \begin{array}{ll}
    \psi_{A,i}(\textbf{r},t^{\prime})=M(\textbf{r})U(t^\prime,t)\left[\psi_{A,i}(\textbf{r},t)+\psi_{B,i}(\textbf{r},t)\right] \\ 
    \psi_{B,i}(\textbf{r},t^{\prime})=[1-M(\textbf{r})]U(t^\prime,t)\left[\psi_{A,i}(\textbf{r},t)+\psi_{B,i}(\textbf{r},t)\right]
\end{array} \right. 
\end{equation}
where $U(t^{\prime},t)$ is the time propagator associated with the
full Hamiltonian including the external
fields. Equation~(\ref{eq:prop_scheme}) defines a recursive
propagation scheme completely equivalent to a time propagation in the
whole space $A\cup B$.

In typical experimental setups, detectors are situated far away from
the sample and electrons overcoming the ionization barrier travel a
long way before being detected. During their journey toward the
detector, and far away from the molecular system, they practically
evolve as free particles driven by an external field.  It seems
therefore a waste of resources to solve the full Schr\"odinger
equation for the traveling electrons while their behavior can be
described analytically. In addition, an ideal detector placed
relatively close to the molecular region would measure the same PES.

From these observations we conclude that we can reduce region $A$ to
the size of the interaction region and assume electrons in $B$ to be
well described by non-interacting Volkov states.  Volkov states are
the exact solution of the Schr\"odinger equation for free electrons in
an oscillating field.  They are plane-waves and are therefore
naturally described in momentum space.  In the velocity gauge the
Volkov time propagator is formally expressed by
\begin{equation}
   U_V(t^{\prime},t) =  \exp\left\{-i\int_{t}^{t^{\prime}} {\rm d}\tau\,\frac{1}{2}\left[\mathbf{p}- \frac{\mathbf{A}(\tau)}{c} \right]^2\right\} \,,
\end{equation}
where the time-ordering operator is omitted for brevity and
$\mathbf{A}(\tau)$ is the vector potential. This is equivalent to the use of a strong-field 
approximation in the outer region in the same
spirit of the Lewenstein model~\cite{Lewenstein:1994ki}.

In summary, the method we propose
consists in solving numerically the real-space TDDFT equations
in $A$ and analytically propagating the wavefunctions residing in $B$
in momentum space. In this setup region $C$ acts as a communication
layer between functions in $A$ and $B$.  Under this prescription, and
by handling $B$-functions in momentum space, Eq.~(\ref{eq:prop_scheme})
becomes
\begin{equation}\label{eq:full_mask_method}  
  \left\{ 
  \begin{array}{ll}
    \psi_{A,i}(\textbf{r},t^{\prime}) = \eta_{A,i}(\textbf{r},t^{\prime}) + \eta_{B,i}(\textbf{r},t^{\prime})\\ 
    \tilde{\psi}_{B,i}(\textbf{p},t^{\prime})= \tilde{\xi}_{A,i}(\textbf{p},t^{\prime}) + \tilde{\xi}_{B,i}(\textbf{p},t^{\prime})
\end{array} \right. 
\end{equation}
with 

\begin{subequations}
\begin{align}
  \eta_{A,i}(\textbf{r},t^{\prime}) =& M(\textbf{r})U(t^\prime,t)\psi_{A,i}(\textbf{r},t) \label{eq:ba_abs_bound}\\
  \eta_{B,i}(\textbf{r},t^{\prime}) =& M(\textbf{r}) \int \frac{{\rm d}\mathbf{p}\, e^{i \mathbf{p}\cdot \mathbf{r}} }{(2 \pi)^{\frac{d}{2}}} U_V(t^{\prime},t)  \tilde{\psi}_{B,i}(\textbf{p},t)\label{eq:back_action}\\
  \tilde{\xi}_{A,i}(\textbf{p},t^{\prime}) =&  \int \frac{{\rm d}\mathbf{r}\, e^{-i \mathbf{p}\cdot \mathbf{r}} }{(2 \pi)^{\frac{d}{2}}}  \nonumber\\
                                                &\times [1-M(\textbf{r})] U(t^{\prime},t)  \psi_{A,i}(\textbf{r},t) \\
  \tilde{\xi}_{B,i}(\textbf{p},t^{\prime}) =&  U_V(t^{\prime},t)  \tilde{\psi}_{B,i}(\textbf{p},t) \nonumber \\
      \label{eq:ba_correction_B}                                          & -\int \frac{{\rm d}\mathbf{r}\, e^{-i \mathbf{p}\cdot \mathbf{r}} }{(2 \pi)^{\frac{d}{2}}} \eta_{B,i}(\textbf{r},t^{\prime})\,.
\end{align}
\end{subequations}
At each time step the orbital $\psi_{A,i}$ is evolved under the mask
function and stored in $\eta_{A,i}$, forcing $\eta_{A,i}$ to be
localized in $A$. At the same time, the components of $\psi_{A,i}$
escaping from $A$ are collected in momentum space by
$\tilde{\xi}_{A,i}$. We then add to $\tilde{\xi}_{A,i}$ the contribution
of the wavefunctions already present in $B$ at time $t$ by summing up
$ U_V \tilde{\psi}_{B,i}$. In order to allow electrons to come back
from $B$ to $A$ we include $\eta_{B,i}$ in $A$ and correct the
function in $B$ by removing its Fourier components [second term in
Eq.~(\ref{eq:ba_correction_B})].

One of the advantages of Eq.~(\ref{eq:full_mask_method}) is that all the
spatial integrals present in $\eta_{B,i}(\textbf{r},t^{\prime})$ and
$\tilde{\xi}_{B,i}(\textbf{p},t^{\prime})$ are performed on functions
localized in $C$. Therefore, integrals over the whole space are evaluated
at the cost of an integration on the much smaller buffer region $C$
that can be easily evaluated by fast Fourier transform
algorithms. Similar considerations hold for integrals in momentum
space under the assumption that $B$-functions
$\tilde{\psi}_{B,i}(\mathbf{p},t)$ are localized in momentum. When
region $A$ is discretized on a grid, in order to avoid wavefunction
wrapping at the boundaries and preserve numerical stability,
additional care must be taken.  In our implementation, numerical
stability is addressed by the use of non-uniform Fourier transforms
(see details in Appendix~\ref{sec:numerical}).

There are situations were the electron flow from $B$ to $A$ is
negligible. This is the case, for instance, when $A$ is large enough
to contain the whole wavefunctions at the time when the external field
has been switched off. A propagation at later times will see
photoelectrons flowing mainly from $A$ to $B$.  In this situation,
$\eta_{B,i}$ and the corresponding correction term in
$\tilde{\xi}_{B,i}$ can be discarded. The evolution scheme of
Eq.~(\ref{eq:full_mask_method}) is thus simplified and becomes
\begin{equation}\label{eq:mask_method}  
  \left\{ 
  \begin{array}{ll}
    \psi_{A,i}(\textbf{r},t^{\prime})=\eta_{A,i}(\textbf{r},t^{\prime}) \\ 
    \tilde{\psi}_{B,i}(\textbf{p},t^{\prime})=\tilde{\xi}_{A,i}(\textbf{p},t^{\prime}) + U_V(t^{\prime},t)  \tilde{\psi}_{B,i}(\textbf{p},t)
\end{array} \right. 
\end{equation}
In the folowing we will refer to Eq.~(\ref{eq:mask_method}) as the
``mask method'' (MM), and to Eq.~(\ref{eq:full_mask_method}) as the
``full mask method'' (FMM).  We note here again that, being single-particle 
propagations schemes, both MM and FMM are not restricted to
TDDFT and can be applied to other effective single-particle
theories. As a matter of fact an approach similar to
Eq.~(\ref{eq:mask_method}) has already been employed in the
propagation of the TDSE equations for atomic
systems~\cite{Chelkowski:1998ec,Grobe:1999vy,Tong:2006wv}, 
and in TDDFT for one-dimensional models of metal surfaces~\cite{Varsano:2004}.
We also note that the implementation of absorbing boundaries trough a mask function
as done in Eq.~(\ref{eq:ba_abs_bound}) can be cast 
in terms of an additional imaginary potential (exterior complex scaling) 
in the Schr\"odinger equation. Such approach is commonly used in quantum optics.

Within the MM, the evolution in $A$ is completely unaffected by the
wavefunctions in $B$ and ionized electrons are treated uniquely in
momentum space. Compared with the FMM, the MM is numerically more
stable as it is not affected by boundary wrapping. In order to achieve
the conditions where Eq.~(\ref{eq:mask_method}) is valid may require,
however, large simulation boxes. Moreover as the mask function never
absorbs perfectly the electrons, spurious reflections may
appear. Suppression of such artifacts requires a further enlargement
of the buffer region. With FMM spurious reflections are almost
negligible.

Choosing between MM and FMM implies a tradeoff between computational
complexity and numerical stability that strongly depends on the
ionization dynamics of the process under study.  In what follows, we
will illustrate the differences and devise a prescription to help the
choice of the most suitable method in each specific case.



\section{Applications}
\label{sec:applications}

In this section we present a few numerical applications of the schemes
previously derived.  In all calculations the boundary between
$A$ and $B$ regions is chosen as a $d$-dimensional sphere
implemented by the mask function:
\begin{equation}\label{eq:mask_function}
  M(\mathbf{r})=
  \left\{ 
  \begin{array}{ll}
    1 & \mbox{if $r < R_{C}$} \\
    1-\sin^{2}\left( \frac{(r-R_{C})\pi}{2(R_{A}-R_{C})} \right)  & \mbox{if $R_{C}\leq r  \leq R_{A}$} \\ 
    0 & \mbox{if $r > R_{A}$}
\end{array} \right. \,,
\end{equation} 
as shown in Fig.~\ref{fig:mask_function}. Note that numerical studies
(not presented here) revealed a weak dependence of the final results
on the functional shape of the mask.

The time propagation of the orbitals in $A$ is performed with the
enforced time-reversal symmetry evolution operator~\cite{Castro:2004hk}
\begin{equation}
  U(t+\Delta t,t)= \exp\left(-i \frac{\Delta t}{2} H(t+ \Delta t)\right)\exp\left(-i \frac{\Delta t}{2} H(t)\right)\, , 
\end{equation} 
where $H$ is the full KS Hamiltonian, and the coupling with the
external field is expressed in the velocity gauge.

The first system we will study is hydrogen.  In spite of being a one-electron
system, it is a seemingly trivial case that has been and still is
under thorough theoretical
investigation~\cite{Burenkov:2010js,GrumGrzhimailo:2010up,Ivanov:2011ev,Zhou:2011jm,Pullen:2011iu,Zheng:2012fw,Tao:2012ev}.
Clearly we do not need TDDFT to study hydrogen and our numerical
results are obtained by propagating the wavefunction with a
non-interacting Hamiltonian.  The interest in this case is focused on
the numerical performance of different mask methods as hydrogen provides a
useful benchmark.

The full TDDFT calculations performed for molecular nitrogen, carbon
monoxide, and benzene are later presented in Sect.~\ref{sub:N2
  infrared pulse} and Sect.~\ref{sub:HeI_PAD} respectively. In these
cases, norm-conserving Troullier-Martins pseudopotentials and the
exchange-correlation LB94 potential~\cite{vanLeeuwen:1994bh} (that has
the correct asymptotic limit for molecular systems) are employed.
Finally, in all calculations the starting electronic
structure of the molecules is calculated in the Born-Oppenheimer
approximation at the experimental equilibrium geometry and the time
evolution is performed with fixed ions.

\subsection{Photoelectron spectrum of hydrogen} 
\label{sub:multi_photon_ionization_1d}

As first example we study multi-photon ionization of a one-dimensional
soft-core hydrogen atom, initially in the ground state, and exposed to
a $\lambda=532$~nm ($\omega=0.0856$~a.u.) linearly polarized laser
pulse with peak intensity $I=1.38\times 10^{13}$~${\rm W}/{\rm cm}^2$,
of the form
\begin{equation}\label{eq:pulse}
  A(t)= A_0 f(t) \cos(\omega t)
\end{equation}
where $f(t)$ is a trapezoidal envelope function of 14 optical cycles 
with two-cycle linear ramps, constant for 10 cycles, and with
$A_0=31.7$~a.u. Here $A(t)$ is the vector potential in units of the speed of
light $c$. A soft-Coulomb potential $ V(x)=-1/\sqrt{2+x^{2}}$ is
employed to model the electron-ion interaction.  
We propagate the electronic wavefunction in time and then
compare the energy-resolved ionization probability obtained from
different schemes. Along with MM, FMM, and SPM we present results for
direct evaluation of PES from Eq.~(\ref{eq:pes_tddft_e}). In this method the spectrum
is obtained by directly Fourier transforming the wavefunction in
region B. Since the analysis is conducted without perturbing the evolution of the 
wavefunction we will refer to it as the ``passive method'' (PM). 
This method requires the knowledge of the whole wavefunction after
the pulse has been switched off, and since a considerable part of the
wave-packet is far away from the core (for the present case
a box of 500~a.u. radius is needed for 18 optical cycles), it is
viable only for one-dimensional calculations. Nevertheless it is
important as it constitutes the limiting case for both MM and FMM.
\begin{figure}
  \includegraphics[width=8.6cm,keepaspectratio]{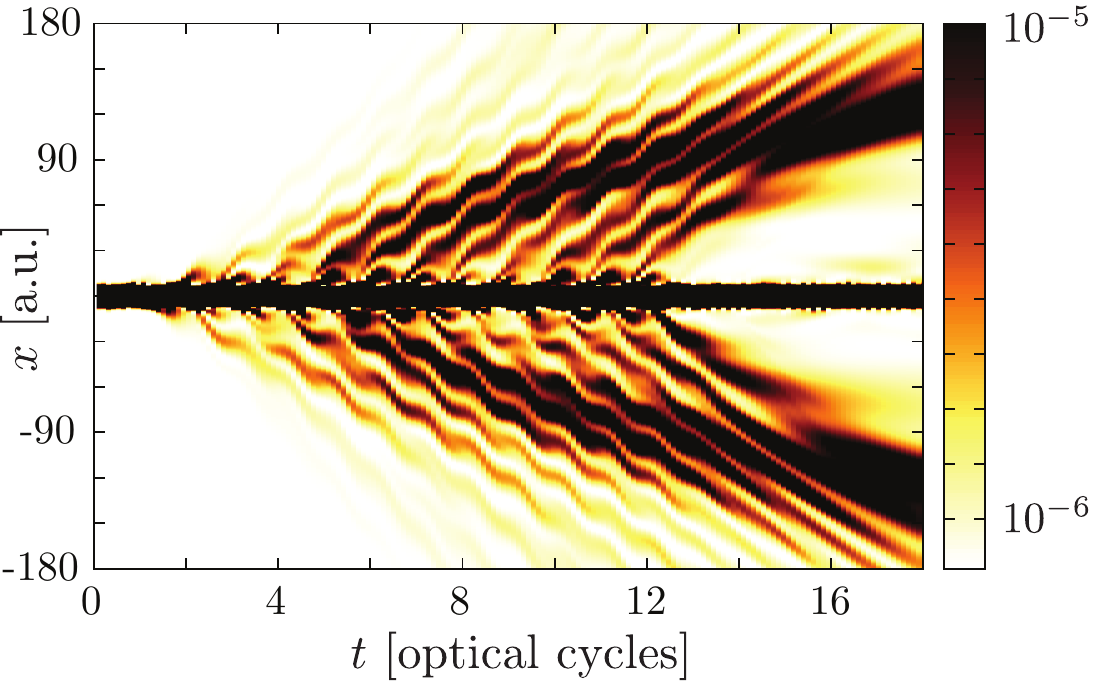}
  \caption{(Color online) Evolution of the electronic density as a function of time
    $\rho(x,t)=|\psi(x,t)|^{2}$ for the one-dimensional soft-core
    hydrogen model. The laser pulse has angular frequency
    $\omega=0.0856$~a.u., intensity $I=1.38\times 10^{13}$~${\rm
      W}/{\rm cm}^2$, and a trapezoidal envelope with 2 optical cycle
    linear ramp (one optical cycle~$= 1.774$~fs) and 10 cycles constant
    center.}
  \label{fig:evo_multi_1d}
\end{figure}

In Fig.~\ref{fig:evo_multi_1d}, a color plot of the evolution of the
electronic density as a function of time is shown. The electronic
wavefunction splits into sub-packets generated at each laser cycle
(one optical cycle~$=2 \pi/\omega= 1.774$~fs). These
wavepackets evolve in bundles and their slope correspond to a certain 
average momentum. 
ATI peaks are then formed by the build up of interfering
wavepackets periodically emitted in the laser field and leading to a given 
final momentum~\cite{Schwengelbeck:1994js}.

\begin{figure}
  \includegraphics[width=8.6cm,keepaspectratio]{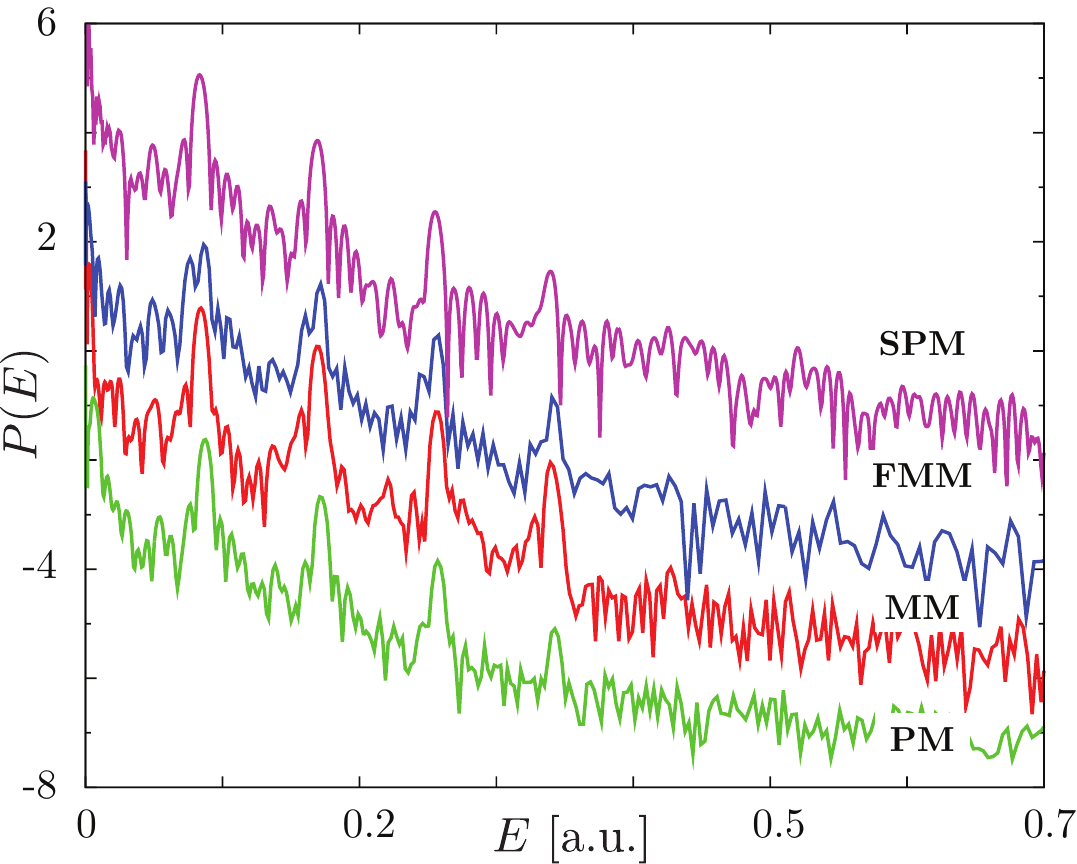}
  \caption{(Color online) Energy resolved photoelectron probability $P(E)$
    (logarithmic scale) calculated with different approaches. The
    spectra are shifted by multiplying a constant factor for easy
    comparison. From bottom to top: passive method
    Eq.~(\ref{eq:pes_tddft_e}) in green, mask method
    Eq.~(\ref{eq:mask_method}) in red, full mask method
    Eq.~(\ref{eq:full_mask_method}) in blue, and sampling point method
    Eq.~(\ref{eq:sampling_point}) in purple. Laser parameters are the same as
    in Fig.~\ref{fig:evo_multi_1d}.}
  \label{fig:pes_multi_1d}
\end{figure}

From Fig.~\ref{fig:evo_multi_1d} is it possible to see that electrons
may be considered as escaped ``already'' at 30~a.u. away from the
center. We set therefore $R_A=30$~a.u. and calculate energy-resolved
PES with the PM. As we can see from Fig.~\ref{fig:pes_multi_1d} the
spectrum presents several peaks at integer multiples of $\omega$
following $E=s \omega -I_P - U_P$ with
$U_P=A_0^2/4c^2=0.0133$~a.u. being the ponderomotive energy,
$I_P=0.5$~a.u. the ionization potential, and $s$ the number of
absorbed photons. In this case the minimum number of photons needed to
exceed the ionization threshold is $s=6$. Of course, the spectrum is
only in qualitative agreement with three-dimensional
calculations~\cite{Schafer:1990gga} as expected from a one-dimensional
soft-core
model~\cite{Kastner:2010gy,Gordon:2005gv,Schwengelbeck:1994js}.

PES calculated from MM, FMM, and SPM all agree as reported in
Fig.~\ref{fig:pes_multi_1d}. Numerical calculations were performed
until convergence was achieved, leading to a grid with spacing $\Delta
R=0.4$~a.u. and box sizes depending on the method. For MM we employed
a simulation box of $R_A=70$~a.u. and set the buffer region at
$R_C=30$~a.u. In order to have energy resolution comparable with PM we
used padding factors (see Appendix~\ref{sec:numerical}) $ P=P_{N}=4$
and the total simulation time was $T=18$ optical cycles. For FMM a smaller box of
$R_A=40$~a.u. with $R_C=30$~a.u. is needed to converge results, and
$P=8$, $P_{N}=2$ were needed to preserve numerical stability for
$T=18$ optical cycles. For SPM two sampling points at $r_{S}=-500,
500$~a.u. were needed to get converged results with a box of 550~a.u.,
and a complex absorber\cite{Kosloff:1986et,Jhala:2010cr} at
49~a.u. from the boundaries of the box. In addition, a total time of
$T=74$ optical cycles was required to collect all the wave packets. The need for
such a huge box resides on the working conditions of SPM. In order to
avoid spurious effects, the sampling points must be set at a distance
such that the density front arrives after the external field has been
switched off. Therefore the longer the pulse the further away the
sampling points must be set. For these laser parameters one could rank
each method according to increasing numerical cost starting from MM,
followed by FMM, PM, and SPM.

As a second example we study the ionization of this one
dimensional hydrogen atom by an ultra-short intense
infrared laser. We employ a single two-cycle pulse of wavelength
$\lambda=800$~nm ($\omega=0.057$~a.u.), intensity $I=2.5\times
10^{14}$~${\rm W}/{\rm cm}^2$, and envelope
\begin{equation}\label{eq:pulse_sin2}
  f(t)=
  \left\{ 
  \begin{array}{ll}
    -\sin(\omega t/2N_{c})^{2} &\mbox{if $0\leq t  \leq 2 \pi N_{c}/\omega$} \\ 
    0  & \mbox{if $t > 2 \pi N_{c}/\omega$}
\end{array} \right. 
\end{equation}
with $N_{c}=2$ and $A_0=225.8$~a.u.
\begin{figure}
  \includegraphics[width=8.6cm,keepaspectratio]{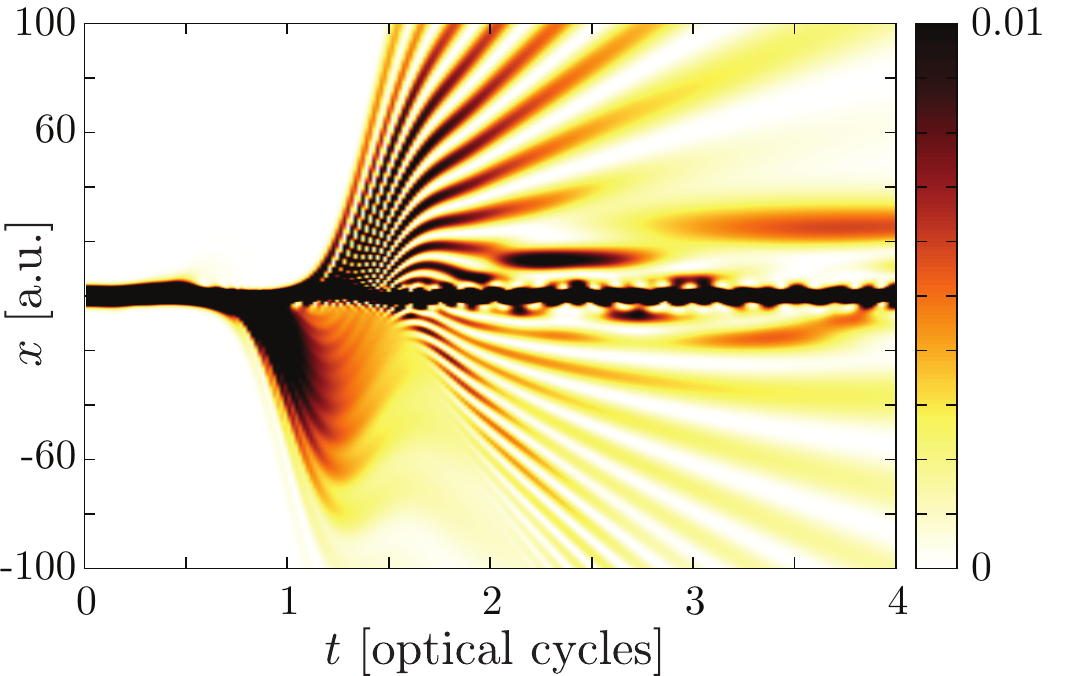}
  \caption{(Color online) Density evolution $\rho(x,t)$ for a one-dimensional
    soft-core hydrogen and a two-cycle $\sin^2$ laser pulse with
    angular frequency $\omega=0.057$~a.u., and intensity $I=2.5\times
    10^{14}$~${\rm W}/{\rm cm}^2$. Here one optical cycle~$=2.66$~fs.  }
  \label{fig:evo_short_1d_pa}
\end{figure}

Due to the laser strength and long wavelength, the electron evolution
shown Fig.~\ref{fig:evo_short_1d_pa} is quite different from the one
presented before. Electrons ejected from the core are driven by the
laser and follow wide trajectories before returning to the parent
ion. Such trajectories can be understood in the context of the
semiclassical model~\cite{Corkum:1993ee} where released electrons move
as a free particle in a time-dependent field with a maximum
oscillation amplitude of $x_0=2A_0/\omega c=57.8$~a.u. Electrons
ejected near a maximum of the electric field $\epsilon(t)=-\partial
A(t)/\partial t$ are the ones gaining the most kinetic energy and are
therefore responsible for the fast emerging electrons after
rescattering with the core.
\begin{figure}
  \includegraphics[width=8.6cm,keepaspectratio]{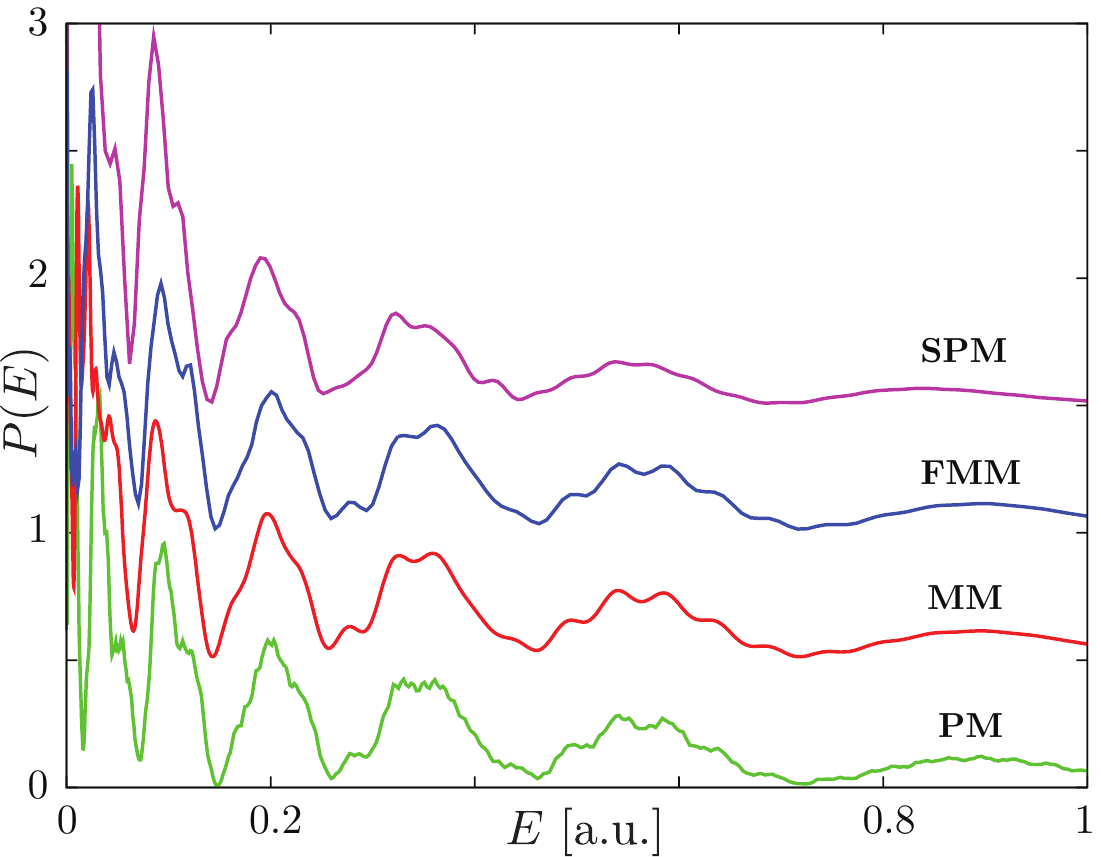}
  \caption{(Color online) Energy-resolved photoelectron $P(E)$ yield from different
    approaches. Spectra are shifted by a constant factor. Order and color coding
    is the same as in Fig.~\ref{fig:pes_multi_1d}. 
    Laser parameters are described in the caption of Fig.~\ref{fig:evo_short_1d_pa}. }
  \label{fig:pes_short_1d}
\end{figure}

In Fig.~\ref{fig:pes_short_1d} we show the energy-resolved PES for
different methods. Here the spectra appear to be very far from any
ATI structure due to short duration of the laser pulse and is
characterized by some irregular maxima and
minima~\cite{Burenkov:2010js}. The characteristic features of the
ionization dynamics is strongly dependent on the detailed shape of the
pulse as one can easily imagine by inspecting the asymmetry in the
electron ejection from Fig.~\ref{fig:evo_short_1d_pa}. Due to these
dynamics, a dramatic carrier envelope phase dependence for such
short pulses is expected.

All the different methods result in similar spectra but with different
parameters. In PM we set $R_{A}=50$~a.u. and a box of radius
$R=700$~a.u. is needed to contain the wave function after $T=4$
optical cycles (one optical cycle~$=2.66$ fs). For MM $R_{A}=200$~a.u.,
$R_C=40$~a.u., and the padding factors are $P=2$, $P_{N}=4$. Here the
value for $R_{A}$ is dictated by the width of the buffer region which
needs to be wide enough to prevent spurious reflections.
\begin{figure}
  \includegraphics[width=8.6cm,keepaspectratio]{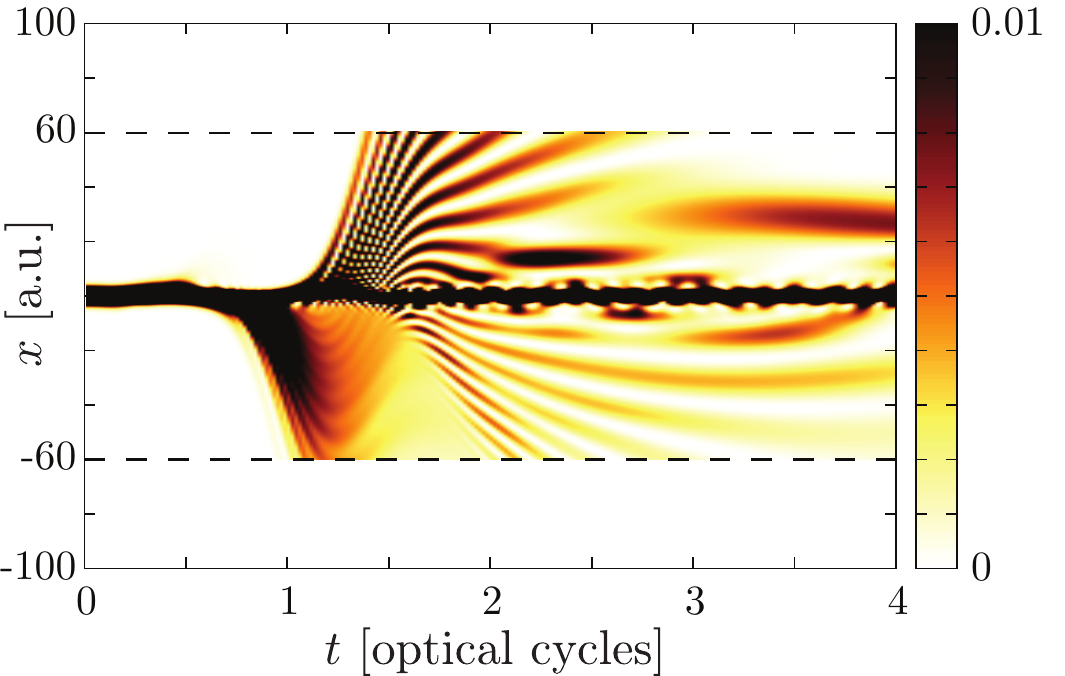}
  \caption{(Color online) FMM density evolution $|\psi_A(x,t)|^2$ of a
    one-dimensional soft-core hydrogen. Laser parameters are the same
    as in Fig.~\ref{fig:evo_short_1d_pa}. Dashed lines indicate the
    edges of the simulation box.}
  \label{fig:evo_short_1d_ba}
\end{figure}
A considerably smaller box is needed for FMM, where $R_A=60$~a.u.,
$R_C=40$~a.u., $P=4$, and $P_{N}=2$. In this case one can reconstruct
the total density in $A$ by evaluating $|\psi_{A}(x,t)|^2$ via
Eq.~(\ref{eq:full_mask_method}) and compare it to the exact
evolution. As one can see in Fig.~\ref{fig:evo_short_1d_ba} the
reconstructed density displays a behavior remarkably similar to the
exact one of Fig.~\ref{fig:evo_short_1d_pa} but with a considerably
reduced computational cost. SMP requires sampling points at
$r_{S}=-130,130$~a.u. in a box of radius $R=200$~a.u. with 49
a.u. wide complex adsorbers, and for a total time of $T=7$ optical cycles.

The possibility to use relatively small simulation boxes is especially
important for three-dimensional calculations where the computational
cost scales with the third power of the box size. Both mask methods
are practicable options for 3D simulations and the advantage of using
FMM with respect to MM is driven by the electron dynamics. For long
laser pulses MM appears to be more stable and it is a better choice
than FMM, while for short pulses with large electron oscillations FMM
can be more performant. SPM is a viable option for short pulses and
small values for the oscillations.

As a last example, we present ATI of a real three-dimensional hydrogen atom
subject to a long infrared pulse. We employ a laser linearly polarized
along the $x$-axis with wavelength $\lambda=800$~nm, intensity
$I=5\times 10^{13}$~${\rm W}/{\rm cm}^2$, pulse shape of the
form~(\ref{eq:pulse_sin2}) with $N_c=20$ and $A_0=91.3$~a.u. Due to to
the pulse length, MM appears to be the most appropriate choice in this
case. In the calculation $R_{A}=60$~a.u., $R_{C}=50$~a.u., $P=1$ and
$P_N=8$.
\begin{figure}
  \includegraphics[width=8.6cm,keepaspectratio]{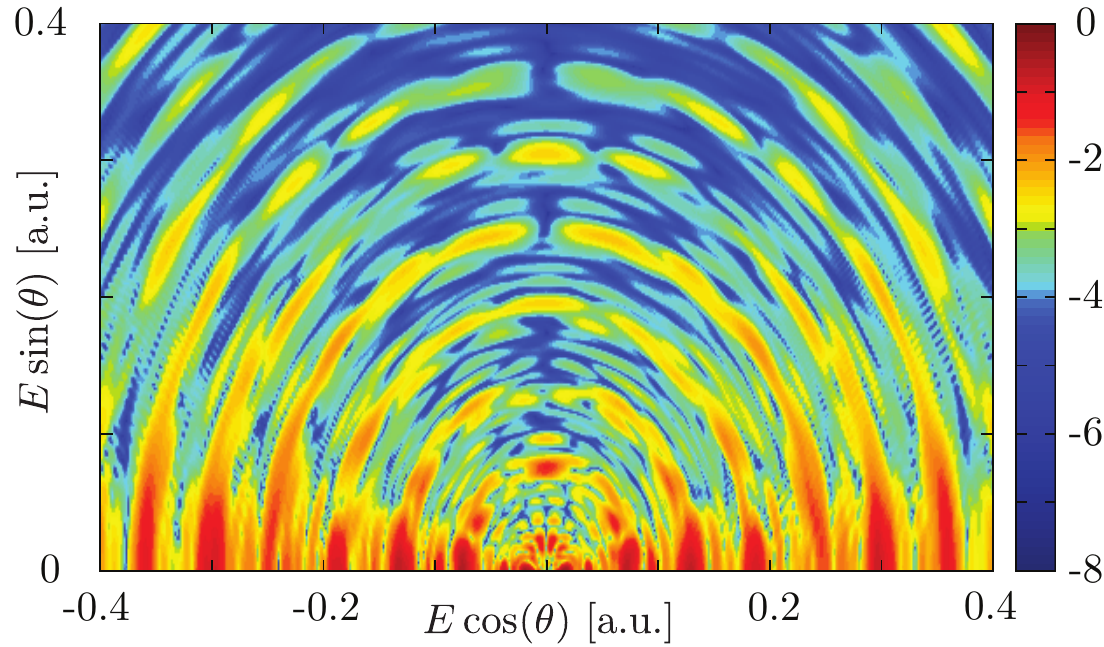}
  \caption{(Color online) Photoelectron angular distribution $P(E,\theta)$ (logarithmic scale) of hydrogen 
  for a 20 cycle $\sin^2$ laser pulse of wavelength $\lambda=800$~nm, 
  and intensity $I=5\times 10^{13}$~${\rm W}/{\rm cm}^2$ polarized 
  along the $x$ axis.}
  \label{fig:pad_h3d}
\end{figure}

In Fig.~\ref{fig:pad_h3d} we show a high-resolution density plot of
the PAD $P(E,\theta)$ defined in Eq.~(\ref{eq:pad}). The radial distance
denotes the photoelectron energy while the angle indicates the
direction of emission with respect to the laser polarization. The
color density is plotted in logarithmic scale and represents the
values of $P(E,\theta)$.

The photoelectron energy-angular distribution displays complex
interference patterns. The pattern shape compares favorably with
similar calculations in the
literature~\cite{Arbo:2008jc,Telnov:2009ie,Zhou:2011jm,Tao:2012ev}. It
consists of a series of rings with fine structures. Each ring
represents the angular distribution of the photoelectron ATI
peaks. The spacing of adjacent rings equals the photon energy
$\omega=0.057$~a.u.
Photoelectrons are emitted mainly along the laser polarization, 
and the left-right symmetry of the rings indicates that the
photoelectrons do not present any preferential ejection 
side with respect to the polarization axis. 
The first ring corresponds to the angular distribution of the first ATI
peak. It presents a peculiar nodal pattern that is induced by the
long-range Coulomb potential and is related to the fact that the ATI
peak is determined by one dominant partial wave in the final
state~\cite{Arbo:2006gg}. The number of the stripes equals the angular
momentum quantum number of the dominant partial wave in the final
state plus one~\cite{Arbo:2006gg}. In Fig.~\ref{fig:pad_h3d}, the
first ring contains six stripes and the dominant final state has
angular momentum quantum number of 5. The pattern of the
energy-angular distribution and the stripe number of the first ring
are in good agreement with those in the
literature~\cite{Telnov:2009ie,Zhou:2011jm}. As for the fine
structures, we observe that while the main ring pattern is already
formed in the first half of the pulse, the fine structure builds up
until the end of the pulse. This supports the hypothesis that such
structures are induced by the coherence of the two contributions from
the leading and trailing edges of the pulse
envelope~\cite{Telnov:2009ie}.


\subsection{N$_2$ under a few-cycle infrared laser pulse} 
\label{sub:N2 infrared pulse}

In this section we compare theoretical and experimental angular
resolved photoelectron probabilities for randomly oriented N$_{2}$
molecules. We choose the laser parameters according to
experiment~\cite{GazibegovicBusuladzic:2011cp}, i.e., we employ a $N_{c}=6$
cycle pulse of wavelength $\lambda = 750$ nm ($\omega=0.06$ a.u.),
intensity $I=4.3\times 10^{13}$ ${\rm W/cm}^2$.
A laser shape
\begin{equation}
  {\bf A}(t)=
  \left\{ 
  \begin{array}{ll}
    \frac{{\bf A}_{0}}{2}(1-\cos(\frac{\omega t}{N_{c}})) \sin(\omega t) & \mbox{if $0\leq t \leq 2\pi N_{c} /\omega$}\\ 
    0 & \mbox{if $t >2 \pi N_{c}/\omega$}
\end{array} \right. 
\end{equation}
for the vector potential should lead to an electric field
similar to the one employed in the experiment with zero carrier
envelope phase.
\begin{figure}
  \includegraphics[width=8.6cm,keepaspectratio]{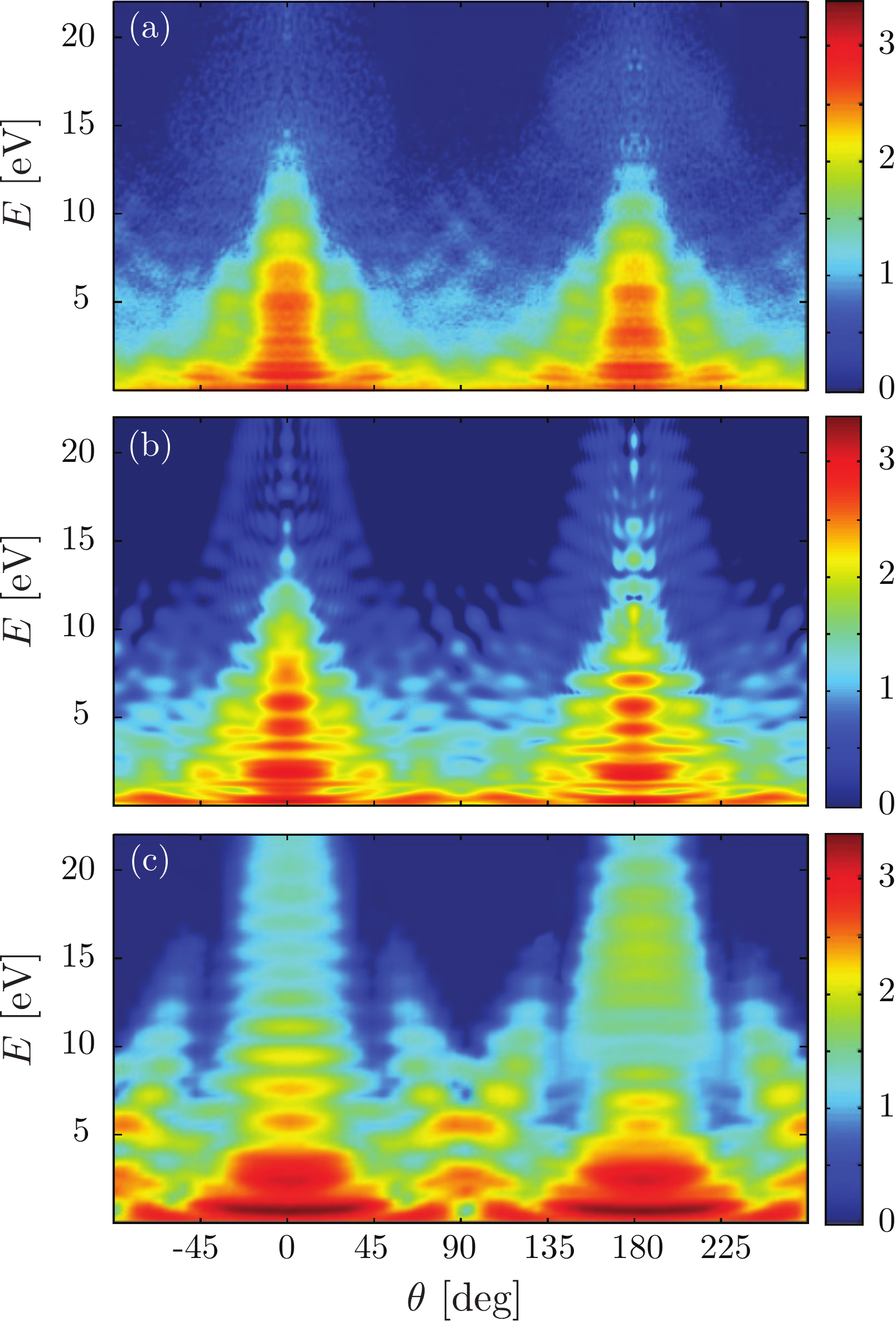}
  \caption{(Color online) Photoelectron angle and energy resolved probability
    $\bar{P}(E,\theta)$ (log scale) in the laboratory frame for
    randomly oriented N$_{2}$ molecules in a 6 cycles infrared laser
    pulse with $\lambda=750$ nm, with intensity $I=4.3\times 10^{13}$
    ${\rm W/cm}^2$. The angle $\theta$ is measured from the laser
    polarization axis. The different panels represent
    $\bar{P}(E,\theta)$ (spanning 3.4 orders of magnitude), from (a)
    experiment, (b) calculated with TDDFT and FMM, and (c) calculated
    with modified molecular strong field approximation. Panels (a) and
    (c) are adapted from
    Ref.~\onlinecite{GazibegovicBusuladzic:2011cp}. }
  \label{fig:pes_n2_average}
\end{figure}

In Fig.~\ref{fig:pes_n2_average}~(a) the experimental photoelectron
probability $\bar{P}(E,\theta)$ is plotted in logarithmic scale as a
function of the energy and the angle with respect to the laser
polarization in the laboratory frame.  Electrons are mainly emitted at
small angles, and, due to the short nature of the pulse electron
emission is asymmetric along the laser polarization axis (at angles
close to $0^{\circ}$ and $180^{\circ}$).

We performed TDDFT calculations for different angles $\theta_{L}$
between the molecular axis and the laser polarization. The molecular
geometry was set at the experimental equilibrium interatomic distance
$R_{0}=2.074$ a.u. The Kohn-Sham wavefunctions were expanded in real
space with spacing $\Delta r=0.38$~a.u. in a simulation box of
$R_{A}=35$~a.u. The photoelectron spectra were calculated with FMM
having $R_{C}=25$ a.u., and padding factors $P=1$, and $P_{N}=4$.
\begin{figure}
  \includegraphics[width=8.6cm,keepaspectratio]{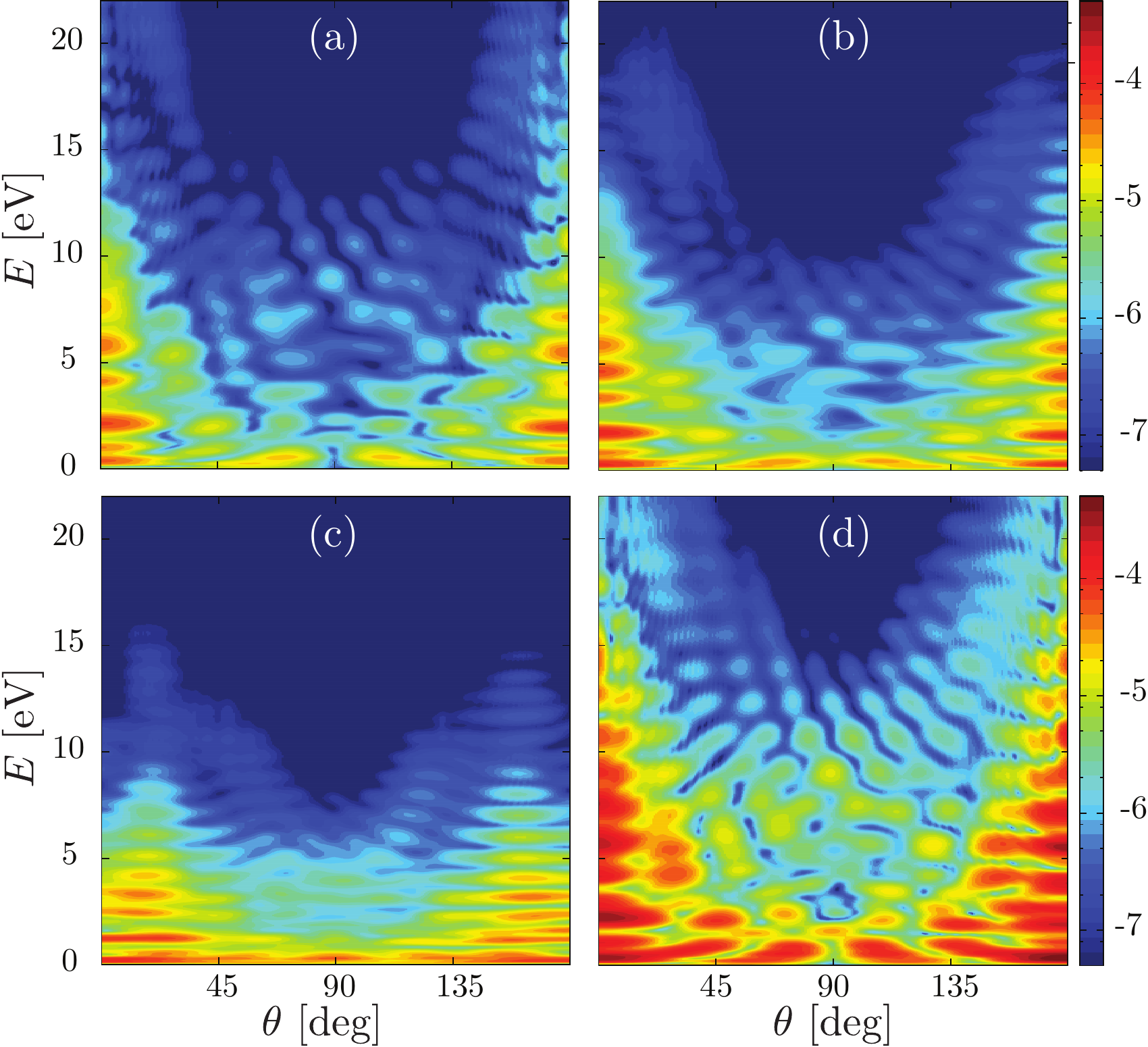}
  \caption{(Color online) Photoelectron angle and energy resolved probability $
    P_{\theta_L}(E,\theta)$ (log scale) for aligned N$_{2}$ molecules
    and different laser polarization directions $\theta_{L}$: (a)
    $\theta_{L}=90^{\circ}$, (b) $\theta_{L}=60^{\circ}$, (c)
    $\theta_{L}=30^{\circ}$, (d) $\theta_{L}=0^{\circ}$. Here
    $\theta_{L}$ is the angle between the laser polarization direction
    and the molecular axis. Laser parameters are the same as in
    Fig.~\ref{fig:pes_n2_average}. }
  \label{fig:pes_n2_panel}
\end{figure}

In Fig.~\ref{fig:pes_n2_panel} the logarithmic ionization probability
$P_{\theta_L}(E,\theta)$ is plotted as a function of energy $E$ and
angle $\theta$ measured from the laser polarization axis for different
values of $\theta_L$.  As the molecular orientation decreases from
$90^{\circ}\leq\theta_{L}\leq 30^{\circ}$ we observe an increasing
suppression of the emission together with a shift of the maximum that
moves away from the laser polarization axis. For
$\theta_{L}=0^{\circ}$ the emission is highly enhanced for all angles
and peaked along the laser direction.  The signature of multi-center
emission interference has been predicted to be particularly marked
when the laser polarization is perpendicular to the molecular
axis~\cite{Busuladzic:2008dv,Busuladzic:2008kn}
(i.e. $\theta_{L}=90^{\circ}$). However, the lowest point in energy of
such a pattern is predicted for $\theta=90^{\circ}$ and $E=\pi^{2}/2
R_{0}^{2}\approx 31$ eV, way above the energy window of observable
photoelectrons produced by our laser. A stronger and longer laser
pulse would be required to extend the rescattering plateau toward
higher energies and therefore to reveal the
pattern~\cite{Okunishi:2009ft}.

In order to reproduce the experimental $\bar{P}(E,\theta)$, an average
over all the possible molecular orientations should be performed. Due
to the axial symmetry of the molecule we can restrict the average to
$0\leq \theta_{L}\leq 90^{\circ}$ and integrate all the contributions
with the proper probability weight~\cite{Wopperer:2010fn}
\begin{equation}\label{eq:n2_average}
  \bar{P}(E,\theta)\propto\int_{0}^{90^{\circ}}{\rm d}\theta_{L}\,\sin\theta_{L}\, P_{\theta_L}(E,\theta)\,.
\end{equation}   

We evaluate Eq.~(\ref{eq:n2_average}) by discretizing the integral in
a sum for $\theta_{L}=0^{\circ}, 30^{\circ}, 60^{\circ}, 90^{\circ}$,
and display the result in Fig.~\ref{fig:pes_n2_average}~(b). Even in
this crude approximation, and without taking into account focal
averaging, the agreement with the experiment is satisfactory and
compares favorably to the molecular strong field approximation shown
in Fig.~\ref{fig:pes_n2_average}~(c). The agreement deteriorates for
low energies where the importance of the Coulomb tail is enhanced
as it is not fully accounted due to limited dimensions of the simulation
box. As a matter of fact the agreement greatly increases for higher
energies.


\subsection{He-(I) PADs for carbon monoxide and benzene} 
\label{sub:HeI_PAD}

In this section we deal with UV ($\omega = 0.78$~a.u.) angular resolved photoemission
triggered by weak lasers. When the external field is weak, non-linear
effects can be discarded and first order perturbation theory can be
applied. In this situation, the momentum resolved PES can be evaluated
by Fermi's golden rule as
\begin{equation}\label{eq:golden_rule}
  P(\mathbf{p})\propto \sum_{i} |\langle \Psi_{f}| \mathbf{A}_{0}\cdot \mathbf{p}| \Psi_{i}\rangle|^{2} \delta(E_{f}-E_{i}-\omega)\, ,
\end{equation}
where $|\Psi_{i}\rangle$ ($|\Psi_{f}\rangle$) is the initial (final)
many-body wavefunction of the system and $\mathbf{A}_{0}$ is the laser
polarization axis.  The difficulty in evaluating
Eq.~(\ref{eq:golden_rule}) lies in the proper treatment of the final
state, which in principle belongs to the continuum of the same
Hamiltonian of $|\Psi_{i}\rangle$. In the simplest approach, it is
approximated by a plane wave (PW).  In this approximation the square root
of the momentum-resolved PES is proportional to the sum of the Fourier
transforms of the initial state wavefunctions
$\tilde{\Psi}_{i}(\mathbf{p})$ corrected by a geometrical factor
$|\mathbf{A}_{0}\cdot \mathbf{p}|$
\begin{equation}\label{eq:pw_full}
  \sqrt{P(\mathbf{p})}\propto \sum_{i}|\mathbf{A}_{0}\cdot \mathbf{p}|\times|\tilde{\Psi}_{i}(\mathbf{p})|\,.
\end{equation}
If photoemission peaks are well resolved in momentum, individual
initial states can be selectively measured. In this case a
correspondence between momentum-resolved PES and electronic states in
reciprocal space can be established.  The range of applicability of
the PW approximation has been discussed in the
literature~\cite{Puschnig:2009ho}. It has been postulated that
Eq.~(\ref{eq:pw_full}) should be valid for (i) $\pi$-conjugated planar
molecules, (ii) constituted by light atoms (H, C, N, O) and for (iii)
photoelectrons emerging with momentum $\mathbf{p}$ almost parallel to
the polarization axis.

Here we restrict ourselves to photoemission from the highest occupied
molecular orbital (HOMO). In this case Eq.~(\ref{eq:pw_full}) becomes
\begin{equation}\label{eq:pw_homo}
  \sqrt{P_{H}(\mathbf{p})}\propto |\mathbf{A}_{0}\cdot \mathbf{p}|\times|\tilde{\Psi}_{H}(\mathbf{p})|\, ,
\end{equation}
the subscript $H$ indicating HOMO-related quantities.  We compare
ab-initio TDDFT and PW PADs evaluated at fixed momentum
$|\mathbf{p}_{H}|=\sqrt{2 E_{H}}$ with $E_H=\omega - E_{B}$ being the
kinetic energy of photoelectrons emitted from the HOMO and $E_{B}$ its
binding energy.

TDDFT numerical calculations are carried out on a grid with spacing
$\Delta r=0.28$~a.u. for benzene and $\Delta r=0.38$~a.u. for CO, in a
simulation box of $R_{A}=30$~a.u.. Photoelectron spectra are
calculated using MM with $R_{C}=20$~a.u. and padding factors $P=1$,
$P_N=8$. A 40 cycles pulse with 8 cycle ramp at the He-(I) frequency
$\omega=0.78$~a.u. and intensity $I=1\times 10^{8}$~${\rm W/cm}^2$ is
employed.

\begin{figure}
  \includegraphics[width=8.6cm,keepaspectratio]{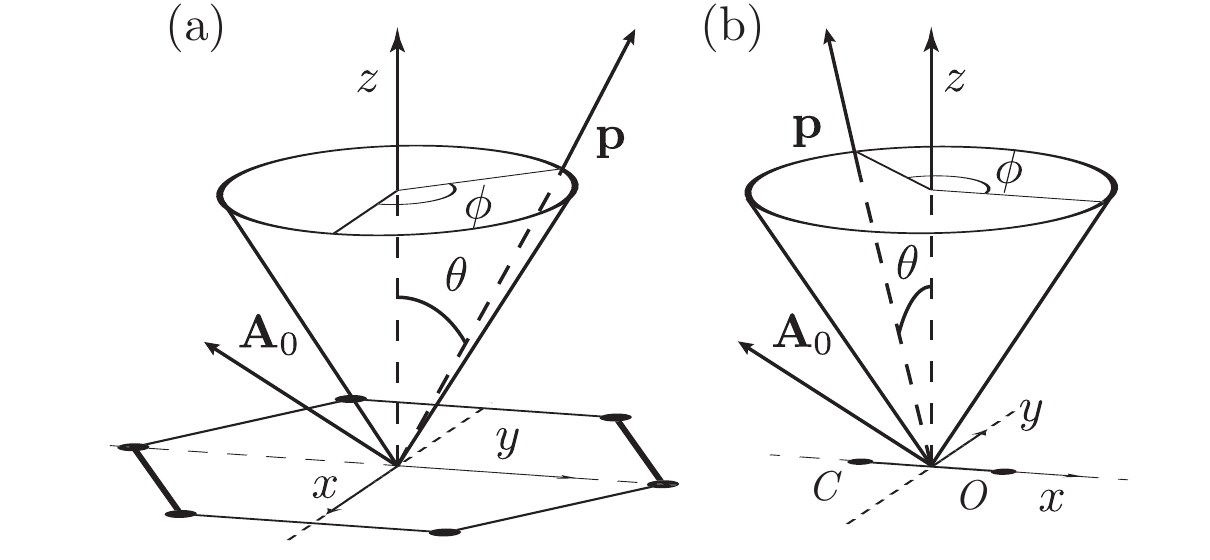}
  \caption{Photoemission geometries for oriented (a) benzene and (b) CO molecules. }
  \label{fig:pad_HeI_geo}
\end{figure}

\begin{figure}
  \includegraphics[width=8.6cm,keepaspectratio]{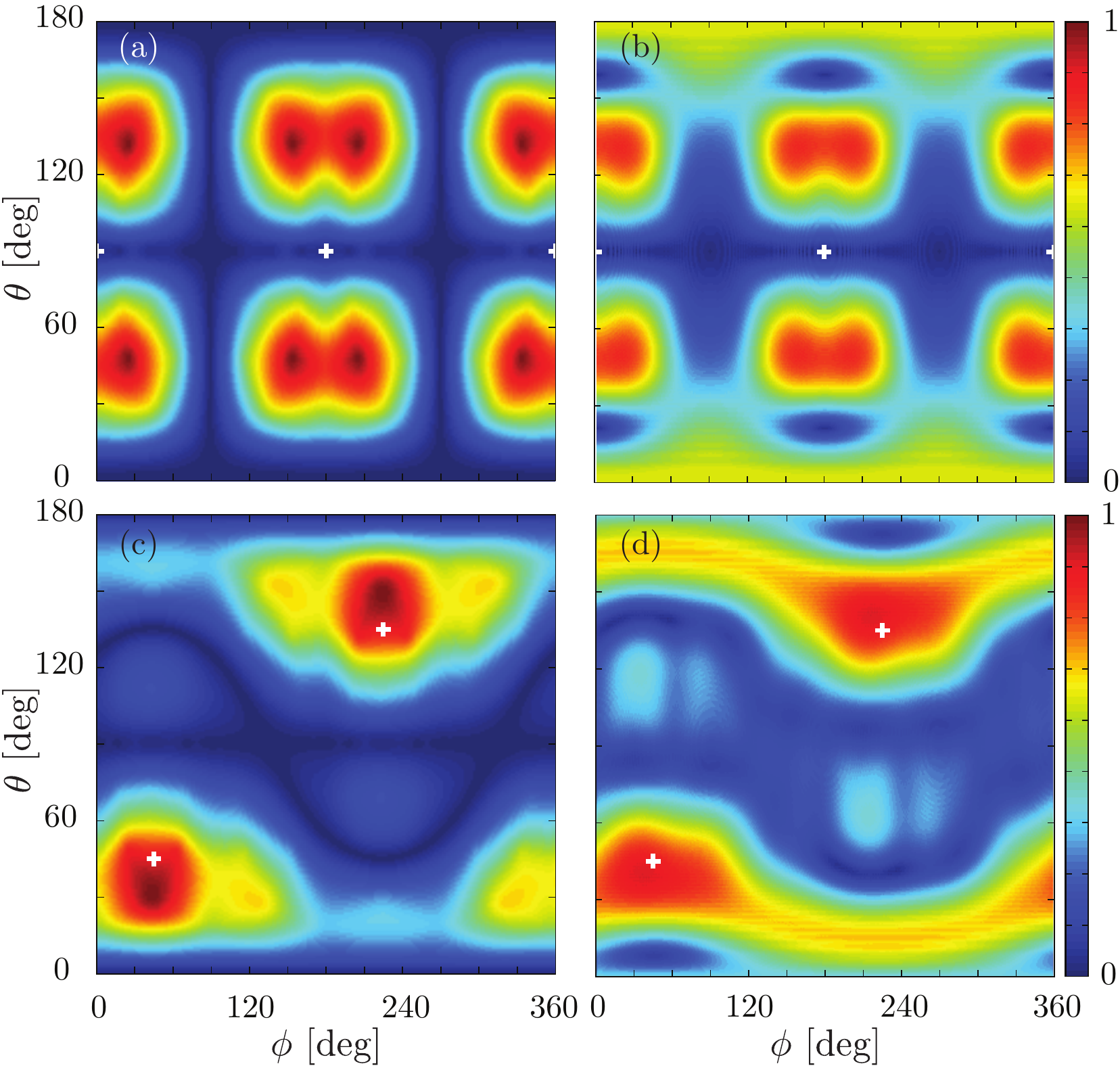}
  \caption{(Color online) He-(I) PADs for aligned benzene molecules.  We compare PADs
    from PW $|\mathbf{A}_{0}\cdot
    \mathbf{p}||\tilde{\Psi}_{H}(\mathbf{p})|$ (left column) and TDDFT
    $\sqrt{P_{H}(\mathbf{p})}$ (right column) on a sphere at constant
    kinetic energy $E_{H}=0.363$~a.u. for different laser
    polarizations $\mathbf{A}_{0}$ (see text for details). Values on
    the sphere are normalized to unity. We used a 40 cycles (8 cycles
    ramp) UV trapezoidal laser pulse with $\lambda=58$~nm
    ($\omega=0.78$~a.u.), and intensity $I=1\times 10^{8}$~${\rm
      W/cm}^2$. In the top row $\mathbf{A}_{0}=\hat{\mathbf{a}}_1=(1,0,0)$, and in the
    bottom row $\mathbf{A}_{0}=\hat{\mathbf{a}}_2=1/\sqrt{3}\times(1,1,1)$. White tics
    indicate the intersection of the laser polarization axis with the
    sphere at constant kinetic energy $E_{H}$. The geometry of the
    photoemission process is indicated in
    Fig.~\ref{fig:pad_HeI_geo}~(a).  }
  \label{fig:pad_benzene_panel}
\end{figure}

We begin presenting the case of benzene since it constitutes the
smallest molecule meeting all the conditions for
Eq.~(\ref{eq:pw_homo}) to be valid.  Results for molecules oriented
according to Fig.~\ref{fig:pad_HeI_geo}~(a), evaluated at
$E_{H}=0.363$ a.u., and two different laser polarizations
$\mathbf{A}_{0}=\hat{\mathbf{a}}_1$,~$\hat{\mathbf{a}}_2$ with
$\hat{\mathbf{a}}_1=(1,0,0)$,
$\hat{\mathbf{a}}_2=1/\sqrt{3}\times(1,1,1)$, are shown in
Fig.~\ref{fig:pad_benzene_panel}.  In the case where the laser is
polarized along the $x$ axis [see
  Fig.~\ref{fig:pad_benzene_panel}~(b)], PAD presents a four lobes
symmetry separated by three horizontal and two vertical nodal lines.
This structure is reminiscent of the HOMO $\pi$-symmetry with the
nodal line at $\theta=90^{\circ}$ corresponding to the nodes of the
orbital on the $x$-$y$ plane.  Information on the orientation of the molecular plane
could then be inferred from the inspection of this nodal
line in the PAD.  A similar feature can be observed also in the case
of an off-plane polarization as shown in
Fig.~\ref{fig:pad_benzene_panel}~(d). In this case, however, the laser
can also excite $\sigma$-orbitals and the nodal line at
$\theta=90^{\circ}$ is partially washed out.  The other nodal lines
can be understood in term of zeros of the polarization factor
$|\mathbf{A}_{0}\cdot \mathbf{p}|$ and are thus purely geometrical.  A
PW approximation of the photoelectron distribution given by
Eq.~(\ref{eq:pw_homo}) qualitatively reproduce the ab-initio results
as shown in Fig.~\ref{fig:pad_benzene_panel}~(a) and (d). According to
condition (iii) a quantitative agreement is reached only for
directions parallel to the polarization axis.
\begin{figure}
  \includegraphics[width=8.6cm,keepaspectratio]{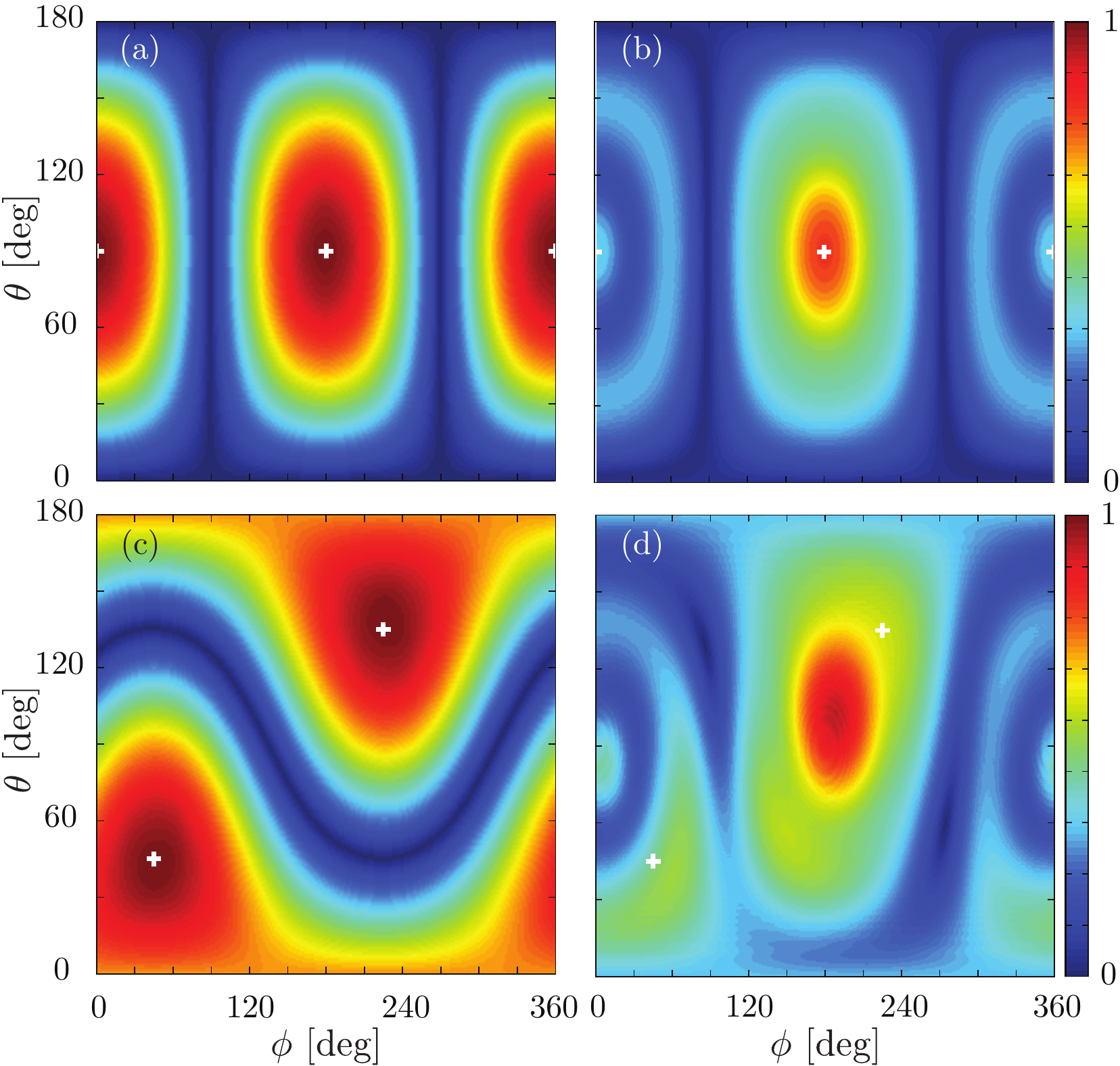}
  \caption{(Color online) He-(I) PADs for aligned CO molecules. Panel ordering and
    laser parameters are the same as in
    Fig.~\ref{fig:pad_benzene_panel}. The molecule is oriented
    according to Fig.~\ref{fig:pad_HeI_geo}~(b) and the photoelectron
    spectra were evaluated on a sphere at $E_{h}=0.261$~a.u.}
  \label{fig:pad_CO_panel}
\end{figure}

A different behavior is expected in the case of CO. Photoelectrons
with kinetic energy of $E_H=0.261$~a.u. are show in
Fig.~\ref{fig:pad_CO_panel}. In this case, condition (i) (i.e. $\pi$-conjugated molecule) is not
fulfilled and a worse agreement between ab-initio and PW calculations
is expected. The quality of the agreement can be assessed by comparing
the left and right columns of Fig.~\ref{fig:pad_CO_panel}. Here, the
weak angular variation of $|\tilde{\Psi}_{H}(\mathbf{p})|$ is
completely masked by the polarization factor $|\mathbf{A}_{0}\cdot
\mathbf{p}|$ [cf. Fig.~\ref{fig:pad_CO_panel}~(a) and (c)].  For this
reason no information on the molecular configuration can be recovered
from a PW model.
 
The situation is qualitatively different for TDDFT as, in this case,
single atom electron emitters are fully accounted for.  Here the nodal
pattern is mainly governed by the polarization factor, but, however,
fingerprints of the molecule electronic configuration can be detected.
For instance, when the laser is polarized along the molecular axis, an
asymmetry of the photoemission maxima can be observed for directions
parallel to $\hat{\mathbf{a}}_1$ [see
  Fig.~\ref{fig:pad_CO_panel}~(b)]. Here the global maximum is peaked
around $(\phi,\theta)=(180^{\circ},90^{\circ})$ corresponding to the
side of the carbon atom on the molecular axis
[cf. Fig.~\ref{fig:pad_HeI_geo}~(b)].  These features can be again
understood in terms of the shape of the HOMO. For CO, in fact, the
HOMO is a $\sigma$ orbital with the electronic charge unevenly
accumulated around the carbon atom. It is therefore natural to expect
photoelectrons to be ejected mainly around the molecular axis and with
higher probability form the side of the carbon atom. This asymmetry is
therefore a property of the electronic configuration of the molecule
and gives information about the molecular orientation itself.  This
behavior appears to be stable upon molecule rotation as can be
observed in the case where the polarization is tilted with respect to
the molecular axis [$\mathbf{A}_{0}=\hat{\mathbf{a}}_1$, see
  Fig.~\ref{fig:pad_CO_panel}~(d)]. Even here the nodal structure is
mainly dictated by the polarization factor.


%
%
\section{Conclusions} 
\label{sec:conclusions}

In this work we studied the problem of photoemission in finite
systems with TDDFT.  We presented a formal derivation of a
photoelectron density functional from a phase-space approach to
photoemission. Such a functional can be directly applied to other
theories based on a single Slater determinant and the derivation could
serve as a base for extensions to more refined models.

We proposed a mixed real- and momentum-space evolution scheme based on
geometrical splitting. In its complete form it allows particles to
seamlessly pass back and forth from a real-space description to a
momentum-space description. The ordinary splitting scheme turns out to
be a special case of this more general method. Furthermore, we
illustrated applications of the method on four physical systems:
hydrogen, molecular nitrogen, carbon monoxide and benzene.

For hydrogen we presented a comparison of the different methods. We
studied ATI peak formation in a one-dimensional model and ATI angular
distributions for a three-dimensional case. The results turned out to
be in good agreement with the literature.  From the comparison, we
derived a prescription to choose the best method based on
a classification of the electron dynamics induced by the external
field.

We investigated angular-resolved photoemission for randomly oriented
N$_2$ molecules in a short intense IR laser pulse.  We illustrated the
results for four different molecular orientations with respect to the
laser polarization. Owing to the symmetry of the problem we were able
to combine the results to account for the random orientation. The
spectrum for randomly oriented molecules is in good agreement with
experimental measurements and is much better than the widely used 
strong field approximations (with one active electron)~\cite{GazibegovicBusuladzic:2011cp}.

We also studied UV angular resolved photoelectron spectra for oriented
carbon monoxide and benzene molecules.  We presented numerical
calculations for two different directions of the laser polarization and
compared with the plane-wave approximation.  We found that the plane-wave
approximation provides a good description for benzene while failing
for CO.  Furthermore, we found evidence that the photoelectron angular
distribution carries important information on molecular orientation.

The successful implementation of photoelectron density functional
presented in this Article paves the road for interesting applications
to many different systems for a wide range of laser parameters.  To
name a few, TDDFT PAD could provide a theoretical tool superior to the
plane-wave and the independent atomic center approximations to retrieve
molecular adsorption orientation information from experiments.
Atto-second pump probe experiments could be simulated ab-initio
accounting for many-body effects but with great computational
advantage with respect to full many-body methods and better physical
description than SAE pictures.

\section{Acknowledgments} 
\label{sec:Acknowledgments}

Special thanks to Lorenzo Stella for many
stimulating discussions and suggestions.  We also wish to acknowledge
useful discussions and comments from Stefan Kurth, Ilya Tokatly, Matteo Gatti
and Franck L\'epine.

Financial support was provided by Spanish (FIS2011-65702-C02-01 and
PIB2010US-00652 ), ACI-Promociona (ACI2009-1036), Grupos Consolidados
UPV/EHU del Gobierno Vasco (IT-319-07), and the European Research
Council Advanced Grant DYNamo (ERC-2010-AdG -Proposal
No. 267374). Computational time was granted by i2basque and BSC “Red
Espanola de Supercomputacion”. MALM acknowledges support from
the French ANR (ANR-08-CEXC8-008-01).


\appendix

\section{Overlap integrals} 
\label{sec:overlap_integral}
In this section we describe the details of the inclusion of the Kohn-Sham
one-body density matrix~(\ref{eq:one_dens_TDDFT}) into
Eq.~(\ref{eq:pes_wigner}). The momentum-resolved photoelectron
probability is the sum over all the occupied orbitals of four overlap
integrals $\gamma$
\begin{equation}\label{eq:pes_gammas}
  P(\mathbf{p})=\sum_{i=1}^{\rm occ.}  \gamma_{A,A,i}(\mathbf{p}) + \gamma_{A,B,i}(\mathbf{p}) + \gamma_{B,A,i}(\mathbf{p}) + \gamma_{B,B,i}(\mathbf{p})\,.
\end{equation}
In order to simplify the notation we drop the orbital index $i$ in the
overlap integrals and indicate with $\mathbf{v}=v \hat{\mathbf{v}}$
the vector $\mathbf{v}$ of modulus $v$ and direction
$\hat{\mathbf{v}}$. In addition, we will consider the simple case
where the boundary surface between region $A$ and $B$ is a
$d$-dimensional sphere of radius $R_{A}$.
 
We start by considering the mixed overlap 
\begin{equation}
  \gamma_{AB}(\mathbf{p})=\int_B{\rm d}\mathbf{R}\,\int \frac{{\rm d}\mathbf{s}}{(2\pi)^{\frac{d}{2}}}\, e^{i \mathbf{p}\cdot \mathbf{s}}  \psi_{A}(\mathbf{R}+\frac{\mathbf{s}}{2})\psi_{B}^{*}(\mathbf{R}-\frac{\mathbf{s}}{2})
\end{equation}
where the integration in $B$ is for $R>R_A$
[cf. Fig.~\ref{fig:geometrical_scheme} (a)].  It is convenient to work
in the coordinates $\mathbf{v}=2 \mathbf{R}$ and
$\mathbf{r}=\mathbf{R}+ \mathbf{s}/2$, where the integral takes the
form
\begin{equation}
  \int_{v>2 R_A}{\rm d}\mathbf{v} \int \frac{{\rm d} \mathbf{r}}{(2\pi)^{\frac{d}{2}}}\,e^{i\mathbf{p}\cdot(2 \mathbf{r}- \mathbf{v})} \psi_{A}(\mathbf{r}) \psi_{B}^{*}(\mathbf{v}- \mathbf{r})\,.
\end{equation}
We substitute $\psi_{B}^{*}$ with its Fourier integral representation
\begin{equation}
  \psi_{B}^{*}(\mathbf{u}- \mathbf{r})=\int \frac{{\rm d} \mathbf{k}}{(2\pi)^{\frac{d}{2}}}\, e^{i \mathbf{k}\cdot(\mathbf{u}- \mathbf{r})} \tilde{\psi}_{B}^{*}(\mathbf{k})
\end{equation} 
and after few simple steps we obtain
\begin{equation}\label{eq:gammaAB}
  \int \frac{{\rm d} \mathbf{k}}{(2\pi)^{\frac{d}{2}}}\, \tilde{\psi}_{A} (- 2 \mathbf{p} -\mathbf{k}) \tilde{\psi}_{B}^{*} (\mathbf{k}) \int_{v>2R_A} {\rm d}\mathbf{v}\, e^{-i (\mathbf{k}+ \mathbf{p})\cdot \mathbf{v}}
\end{equation}
where we successfully disentangled the integration over $\mathbf{v}$
in the second integral. The integral on $v>2 R_A$ can be rewritten as an
integral over the whole space, which yields a $d$-dimensional Dirac
delta, minus an integral on $v\leq2 R_A$:
\begin{multline}\label{eq:besselJ}
  \int_{v>2R_A} {\rm d}\mathbf{v}\, e^{-i (\mathbf{k}+ \mathbf{p})\cdot \mathbf{v}}= \\
  (2\pi)^{d}\delta(\mathbf{k}+\mathbf{p}) - (4\pi R_A)^{\frac{d}{2}} \frac{J_{d/2}(2R_A|\mathbf{k}+\mathbf{p}|)}{|\mathbf{k}+\mathbf{p}|^{\frac{d}{2}}} 
\end{multline}
where $J_n(k)$ is a Bessel function of the first kind. The second term
in~(\ref{eq:besselJ}) is a function centered in $- \mathbf{p}$ and
strongly peaked in the region $w=C_d/R_A$ with $C_1=\pi$, $C_2\approx
3.83$, $C_3\approx 4.49$ being the first zeros of the Bessel function
$J_{d/2}(k)$. If the region $w$ is small enough we can consider the
integrand in $\mathbf{k}$ of~(\ref{eq:gammaAB}) constant and factor
out of the integrand $ \tilde{\psi}_{A} (- 2 \mathbf{p} -\mathbf{k})
\tilde{\psi}_{B}^{*} (\mathbf{k})$ evaluated at $\mathbf{k}=
-\mathbf{p}$. It is easy to see that
\begin{equation}
  \int \frac{{\rm d} \mathbf{k}}{(2\pi)^{\frac{d}{2}}}\, (2 R_A)^{\frac{d}{2}} \frac{J_{d/2}(2R_A|\mathbf{k}+\mathbf{p}|)}{|\mathbf{k}+\mathbf{p}|^{\frac{d}{2}}} = 1 
\end{equation}  
and, by plugging~(\ref{eq:besselJ}) in~(\ref{eq:gammaAB}), we have
that $\gamma_{A,B}(\mathbf{p})\approx0$.  By the same reasoning we
should expect $\gamma_{B,A}(\mathbf{p})\approx0$.

We now turn to the terms containing wavefunction on the same
region. In $(\mathbf{v},\mathbf{r})$ coordinates
\begin{equation}
\gamma_{A,A}(\mathbf{p})=    \int_{v>2 R_A}{\rm d}\mathbf{v} \int \frac{{\rm d} \mathbf{r}}{(2\pi)^{\frac{d}{2}}}\,e^{i\mathbf{p}\cdot(2 \mathbf{r}- \mathbf{v})} \psi_{A}(\mathbf{r}) \psi_{A}^{*}(\mathbf{v}- \mathbf{r})
 \,.
\end{equation}
The product of functions localized in $A$ is not negligible only for
$r<R_A$ and $|\mathbf{v}- \mathbf{r}|<R_A$. Since the integral is
carried out for $v>2 R_A$ we can bound $|\mathbf{v}- \mathbf{r}|$ from
below with $R_A|2\hat{\mathbf{v}} - \mathbf{r}/R_A|\geq R_A$. This
leads to $R_A\leq|\mathbf{v}- \mathbf{r}|< R_A$ which is satisfied
only on the boundary of $A$. Being a set of negligible measure we have
$\gamma_{A,A}(\mathbf{p})=0$.

Once again, in $(\mathbf{v},\mathbf{r})$ coordinates
\begin{equation}
\gamma_{B,B}(\mathbf{p})=    \int_{v>2 R_A}{\rm d}\mathbf{v} \int \frac{{\rm d} \mathbf{r}}{(2\pi)^{\frac{d}{2}}}\,e^{i\mathbf{p}\cdot(2 \mathbf{r}- \mathbf{v})} \psi_{B}(\mathbf{r}) \psi_{B}^{*}(\mathbf{v}- \mathbf{r})
\end{equation} 
can be written as 
\begin{multline}
\label{A10}
\gamma_{B,B}(\mathbf{p})= |\psi_B(\mathbf{p})|^{2} - \\
\int_{v<2 R_A}{\rm d}\mathbf{v} \int \frac{{\rm d} \mathbf{r}}{(2\pi)^{\frac{d}{2}}}\,e^{i\mathbf{p}\cdot(2 \mathbf{r}- \mathbf{v})} \psi_{B}(\mathbf{r}) \psi_{B}^{*}(\mathbf{v}- \mathbf{r})
\end{multline} 
where the first integration is in region $A$. Using the localization of $\psi_B$ we see that the integral is non-zero only for $r>R_A$ and $|\mathbf{v}- \mathbf{r}|>R_A$. As the integration is for $v<2R_A$ we have that $R_A\geq|\mathbf{v}- \mathbf{r}|> R_A$ and therefore the double integral in Eq.~(\ref{A10}) is zero.


\section{Numerical stability and Fourier integrals} 
\label{sec:numerical}

A real-space implementation of Eq.~(\ref{eq:full_mask_method})
involves the evaluation of several Fourier integrals. Such integrals
are necessarily substituted by their discrete equivalent, and
therefore discrete Fourier transforms (FT) and fast
Fourier transforms (FFTs) are called into play. However, evolution
methods based on the discrete FT naturally impose periodic boundary
conditions. While this is not presenting any particular issue for MM where FT
are only used to map real-space wavefunctions to momentum space,
it is a source of numerical instability for FMM where the 
wavefunctions are reintroduced in the simulation box.

The problem is well illustrated by the following one-dimensional
example. Imagine a wavepacket freely propagating to an edge of
the simulation box with a certain velocity. In MM, when passing trough
the buffer region, the packet is converted by discrete FT in momentum space and
then analytically evolved as a free particle through the edge of the
box.
In FMM as the wavefunction evolves in momentum spaces it is also transformed back
to real space to account for possible charge returns.
In this case, instead of just disappearing from one edge, by virtue of the discrete FT periodic boundary conditions, 
the same wavepacket will appear from the opposite side.
It can be easily understood how such an undesirable event can create a
feedback leading to an uncontrolled and unphysical build up of the density.

This behavior can be controlled by the use of zero padding. As we
know, the Fourier integrals in Eq.~(\ref{eq:full_mask_method}) involves
functions that are, by construction, zero outside the buffer region
$C$. We can therefore enlarge the integration domain (having radius
$R_{A}$) by a padding factor $P$, set the integrand to zero in the
extended points, obtaining the same result. As a consequence,
a wavepacket propagating toward a boundary edge will have to run an enlarged {\em virtual} box of radius
$\tilde{R}_{A}=R_{A} (2P -1)$ before emerging from the other side. 
In addition, the smallest momentum represented $\Delta \tilde{p}=2\pi/P
R_{A}=\Delta p/P$ in the discretized $\tilde{\psi}_{B,i}(\textbf{p},t^{\prime})$ 
is reduced by a factor $1/P$ while the highest momentum $p_{max}=\pi/\Delta r$ remains
unchanged. The price to pay here is an
increased memory requirement by a factor $P^d$ (where $d$ is the
dimension of the simulation box) and is too high for three-dimensional
calculations.

A possible way to find a better scaling is offered by the use of  {\em
  non-uniform discrete Fourier Transform} and companion fast algorithm
NFFT~\cite{Kunis:2006wg,Keiner:2009js,Keiner:2009js}. NFFT allow for
the possibility to perform Fourier integrals on unstructured sampling
points with, for fixed accuracy, the same arithmetical complexity as
FFT. For a detailed description of the algorithm we refer to the
literature~\cite{Keiner:2009js}. 
The idea is to use the flexibility of NFFT to perform zero padding in 
a convenient way.
Instead of allocating an enlarged
box filled with zeros at equally spaced sample positions, we set only one
point at $R_{A} P_{N}$ (here $P_{N}$ is the NFFT padding factor) and
evaluate the Fourier integral with NFFT. In this way we gain numerical
stability for FMM as long as the wavefunctions are contained in a {\em virtual} box
of $\tilde{R}_{A}=R_{A} (2P_{N} -1)$ at the price of adding a number
of points that scales as $d-1$ with the dimension of the box. If $N^d$
is the number of grid points in the simulation box, in order to perform zero padding
with NFFT one needs to add only $2N^{d-1}$ points.

With this procedure however, not only the smallest momentum $\Delta
\tilde{p}$ is reduced by a factor $1/P_{N}$, but also the highest
momentum $\tilde{p}_{max}=(N/2+1) \Delta \tilde{p}$ is decreased by
the same amount. This turns out to be the limiting factor in the use of NFFT 
to preserve numerical stability with FMM as the enlargement factor $P_{N}$ has an
upper bound that depends on the escaping electron dynamics. In fact,
when we evaluate the back-action term Eq.~(\ref{eq:back_action}), we assume
$\tilde{\psi}_{B,i}(\textbf{p},t^{\prime})$ to be localized in
momentum and, in order to preserve numerical consistency, $P_N$
must be limited by the highest momentum contained in
$\tilde{\psi}_{B,i}$. A combination of ordinary padding and NFFT
padding helps to balance the tradeoff between memory occupancy and
numerical stability.

Finally, in MM zero padding can be used to increase resolution in momentum.


\end{document}